\documentclass[9pt,twoside,a4paper]{arXiv}
\usepackage[]{geometry}
\usepackage[scriptsize,tight]{subfigure}
\usepackage{graphicx}
\usepackage{multirow}
\usepackage{dcolumn}
\usepackage{bm}
\usepackage{color,soul}
\usepackage{hyperref}

\articletype{inv}

\runningtitle{Adsorption of Greenhouse Gases on Carbon Nanobelts}

\runningauthor{Aguiar \textit{et al.}}

\title{Adsorption Behavior of Greenhouse Gases on Carbon Nanobelts: A Semi-Empirical Tight-Binding Approach for Environmental Application}

\author[1]{C.~Aguiar}

\author[1,$\ast$]{I.~Camps}
\correspondingauthoraffiliation[$\ast$]{icamps@unifal-mg.edu.br}

\affil{Laborat\'orio de Modelagem Computacional - \emph{La}Model,
Instituto de Ci\^{e}ncias Exatas - ICEx. Universidade Federal de Alfenas -
UNIFAL-MG, Alfenas, Minas Gerais, Brazil}

\begin{abstract}
This research investigates the adsorption characteristics of carbon nanobelts (CNB) and Möbius carbon nanobelts (MCNB) interacting with various greenhouse gases, including NH\textsubscript{3}, CO\textsubscript{2}, CO, H\textsubscript{2}S, CH\textsubscript{4}, CH\textsubscript{3}OH, NO\textsubscript{2}, NO, and COCl\textsubscript{2}. The study employs semi-empirical tight-binding calculations via xTB software complemented by topological analysis using MULTIWFN software. Comparative analysis reveals MCNB's superior adsorption properties, particularly for specific gases. Notable adsorption energies for MCNB were measured at~-1.595~eV, -0.669~eV, and -0.637~eV for NO, COCl\textsubscript{2}, and NO\textsubscript{2}, respectively, significantly exceeding the corresponding CNB values of~-0.636~eV, -0.449~eV, and~-0.438~eV. The investigation of desorption kinetics demonstrates rapid recovery times (sub-millisecond) for most gas-nanobelt interactions, with the notable exception of the MCNB+NO system, which exhibits persistent bonding. Topological analysis confirms chemisorption mechanisms for NO, COCl\textsubscript{2}, and NO\textsubscript{2} on both nanobelt variants, characterized by complex hybridizations of covalent and non-covalent interactions. Molecular dynamics simulations conducted in both packed configurations and dry air mixtures demonstrate the nanobelts' effective gas--attracting properties, maintaining consistent capture performance across different environmental conditions. These findings establish carbon nanobelts, particularly the Möbius configuration, as promising candidates for greenhouse gas capture technologies, offering potential applications in environmental remediation and climate change mitigation strategies.
\end{abstract}


\begin{document}
\maketitle
\thispagestyle{firststyle}

\section{INTRODUCTION}
\label{Sec:Intro}

The rapid industrialization, improper waste management, and fossil fuel combustion have led to unprecedented levels of environmental pollution~\cite{Lee-Sens.ActuatorsBChem.-255-1788-2018,Manisalidis-Front.PublicHealth-8--2020,Nool-SSM-PopulationHealth-15-100879-2021}. Various toxic gases, including phosgene, hydrogen sulfide, ammonia, nitric oxide, methanol, methane, carbon monoxide, and carbon dioxide, pose severe risks to human health and environmental integrity~\cite{Lee-Sens.ActuatorsBChem.-255-1788-2018}.

Methanol (CH\textsubscript{3}OH), a hazardous industrial solvent, can cause severe illness or death upon prolonged exposure~\cite{Holt-Intern.Med.J.-48-335-2018}. Methane (CH\textsubscript{4}), while less toxic, presents significant hazards due to its high combustibility~\cite{Jo-Tuberc.Respir.Dis.-74-120-2013}. Carbon monoxide (CO), an imperceptible gas produced by incomplete combustion, can cause fatal poisoning before symptoms become apparent~\cite{Buboltz2023,Otterness-Emerg.Med.Pract.-20-1-2018}. Elevated carbon dioxide (CO\textsubscript{2}) levels can induce oxygen displacement, leading to asphyxiation, acidosis, cardiac arrhythmias, and tissue damage~\cite{Schrier-Front.Toxicol.-4--2022}. Phosgene (COCl\textsubscript{2}) directly compromises respiratory epithelium through its high chemical reactivity~\cite{Rendell-Toxicol.Lett.-290-145-2018,Pauluhn-Toxicology-450-152682-2021}.

Hydrogen sulfide (H\textsubscript{2}S), a highly combustible gas, can cause rapid fatal toxicity at high concentrations~\cite{Ng-J.Med.Toxicol.-15-287-2019}. Ammonia (NH\textsubscript{3}) induces severe tissue necrosis through exothermic reactions~\cite{Pangeni-AnnalsofMedicine&Surgery-82-104741-2022}. Nitrogen oxides (NO and NO\textsubscript{2}) can trigger acute respiratory distress syndrome~\cite{Amaducci2023}, with NO\textsubscript{2} further contributing to acid rain and ozone formation through photochemical reactions~\cite{Lee-Sens.ActuatorsBChem.-255-1788-2018,Verma-ACSSensors-8-3320-2023}.

These environmental hazards necessitate robust gas monitoring systems to maintain safe atmospheric conditions~\cite{Nazemi-Sensors-19-1285-2019,Abooali-J.Comput.Electron.-19-1373-2020,Ahmed-R.Soc.OpenSci.-9-220778-2022}. Two-dimensional (2D) materials have emerged as promising gas sensors due to their large surface area, selective adsorption properties~\cite{Calvaresi-J.Mater.Chem.A-2-12123-2014a,Cezar-arXiv-2023}, chemical stability~\cite{Cezar-arXiv-2023,Chang-ACSNano-4-5095-2010}, and superior electrochemical performance~\cite{Holt-Science-312-1034-2006,Cezar-arXiv-2023}. For instance, phosphorene has demonstrated effective detection of NH\textsubscript{3}, SO\textsubscript{2}, NO, and NO\textsubscript{2}~\cite{Safari-Appl.Surf.Sci.-464-153-2019,Tang-Sensors-21-1443-2021}. Wu \emph{et al.}~\cite{Wu-RSCAdvances-14-1445-2024} showed that functionalized arsenene efficiently detects CO, NO, NO\textsubscript{2}, SO\textsubscript{2}, NH\textsubscript{3}, and H\textsubscript{2}S. Additionally, graphene-based materials have shown promise in detecting CO\textsubscript{2}, CH\textsubscript{4}, and N\textsubscript{2}~\cite{Oliveira-Sci.Rep.-12-22393-2022,Li-Mater.Chem.Phys.-301-127602-2023}.

The global greenhouse gas (GHG) crisis poses an urgent environmental threat, with rising levels of CO\textsubscript{2}, CH\textsubscript{4}, and other heat--trapping gases driving global warming and climate change. As GHG concentrations far exceed pre--industrial levels, the Paris Agreement's goal to limit warming below 2~\textsuperscript{o}C demands immediate action. Achieving this target necessitates both substantial emissions reductions and the development of innovative capture technologies~\cite{Delangiz-ArabJGeosci-12-174-2019,Panepinto-IntJEnvironResPublicHealth-18-6767-2021,Green2022,Xu-FrontEnergyRes-10-849490-2022,Sarika2023,Arinushkina2023}.

Carbon nanobelts (CNBs) are emerging as revolutionary nanomaterials, garnering significant attention due to their exceptional structural, electronic, and mechanical properties. Derived from graphene, these materials bridge the gap between molecular and bulk carbon structures, offering unprecedented potential in diverse fields such as electronics, optoelectronics, and energy storage. The ability to precisely engineer CNBs' structure, including length, width, and edge configurations, further enhances their versatility and functionality across various applications~\cite{Madima-EnvironChemLett-18-1169-2020,Fathy-EgyptJChem-64-7029-2021,Bag2021,Rasyotra2024,Thayanithi-PhysSciRev---2024,Harini2024}

Of particular interest are CNBs composed of 4--5--6--8--member rings, which exhibit thermal stability and direct-gap semiconductor properties. With band gaps ranging from 1.12 to 1.25~eV, these structures are ideally suited for advanced electronic and optoelectronic applications~\cite{Mortazavi-JComposSci-7-269-2023}. The development of sophisticated fabrication methods, including lithographic techniques, catalytic cutting, chemical assembly, and epitaxial growth, has enabled the production of CNBs with finely tuned electronic characteristics~\cite{Celis-JPhysApplPhys-49-143001-2016}. Furthermore, the precise control over CNB structural parameters, such as chirality and handedness, allows for the creation of materials with tailored electronic properties, significantly expanding their potential in cutting-edge fields like spintronics and next-generation energy storage systems~\cite{Gu-JAmChemSoc-144-11499-2022}.

Carbon nanobelts (CNBs) possess several advantageous properties that render them promising candidates for greenhouse gas interactions. These properties include high surface area, tunable chemical functionality, and unique electronic characteristics. The ability to modify their edges and surfaces further enhances their gas adsorption and separation capabilities, making them particularly suitable for environmental applications.

The utilization of nanobelts composed of various materials in the fabrication of membranes for water treatment~\cite{Karkooti-JMembraneSci-560-97-2018} and gas capture has proven effective due to their high efficiency and selectivity~\cite{Sonawane_2021,Khraisheh-Membranes-11-995-2021,Ji-JMembraneSci-683-121856-2023,Hedar_2024}. Graphene oxide nanobelts offer extensive surface area and effective barrier properties, enabling efficient contaminant removal and selective gas retention~\cite{Kwon-Nanomaterials-11-757-2021,Choi-ChemEngJ-427-131805-2022}. Among hybrid membrane architectures, graphene oxide nanobelt (GONB)/polymer composites are noteworthy, where the polymer charge effect influences H\textsubscript{2}/CO\textsubscript{2} separation performance. Gas transport properties can be readily tuned through polymer addition, with performance dependent on the quantity, molecular weight, and charge of these polymers~\cite{Ji-JMembraneSci-683-121856-2023}.

In this study, we investigate the interactions between two types of carbon nanobelts and a range of greenhouse gases, including phosgene, nitric oxide, methanol, methane, carbon monoxide, carbon dioxide, hydrogen sulfide, ammonia, and nitrogen dioxide. We employ semiempirical tight-binding theory to elucidate these interactions at the molecular level. Our comprehensive approach encompasses multiple analytical methods: identification of optimal interaction regions, geometry optimization, electronic property calculations, and topological studies. This multifaceted analysis provides a thorough characterization of the CNB-greenhouse gas systems, offering insights into their potential for environmental remediation applications.

\section{MATERIALS AND METHODS}
\label{Sec:Method}

\begin{figure}[htpb]
\centering
\includegraphics[width=14cm]{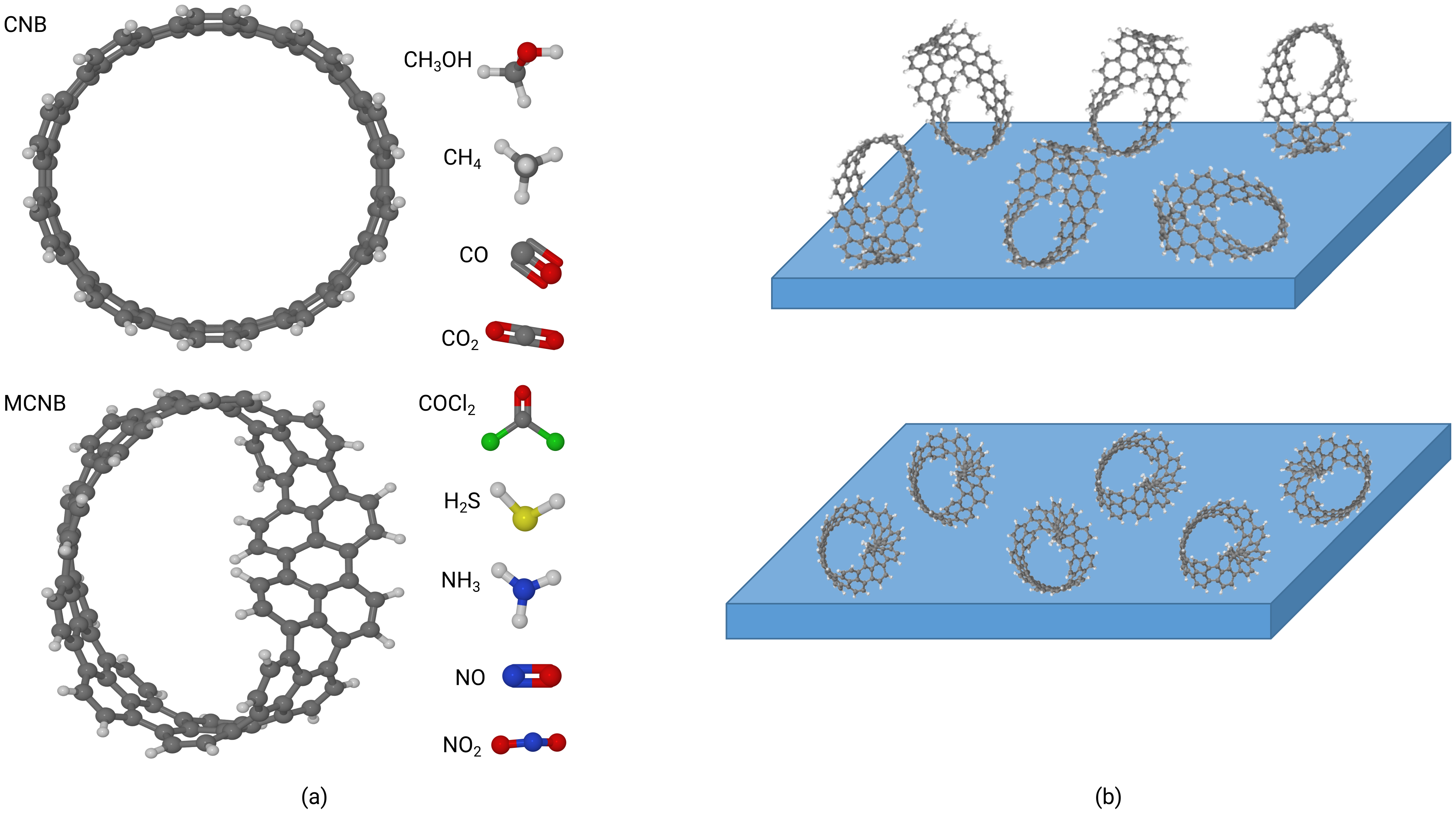}
\caption{(a) Initial structures used. (b) Membrane conformations using the carbon nanobelts.}
\label{Fig:Structures}
\end{figure}

This investigation focused on two distinct carbon-based nanostructures: a standard nanobelt and a Möbius (or twisted) nanobelt. The Virtual NanoLab Atomistix Toolkit software~\cite{VNL} was used to construct the nanobelt, starting with two unit cells of a (10,0) carbon nanosheet replicated ten times along the z-axis. The standard nanobelt was then formed by wrapping this structure through a full 360-degree rotation, removing periodicity, and passivating the edge atoms with hydrogen. For the Möbius variant, an additional 180-degree twist was introduced before wrapping. The study examined the interaction of these nanostructures with nine greenhouse gases: nitrogen-dioxide (NO\textsubscript{2}), hydrogen-sulfide (H\textsubscript{2}S), methanol (CH\textsubscript{3}OH), methane (CH\textsubscript{4}), carbon monoxide (CO), carbon dioxide (CO\textsubscript{2}), phosgene (COCl\textsubscript{2}), ammonia (NH\textsubscript{3}), and nitrogen-oxide (NO). All molecular structures were maintained in a neutral charge state. Visual representations of these structures together with a possible configuration in membrane formation are provided in Figure~\ref{Fig:Structures}.

\begin{figure}[htpb]
\centering
\includegraphics[width=14cm]{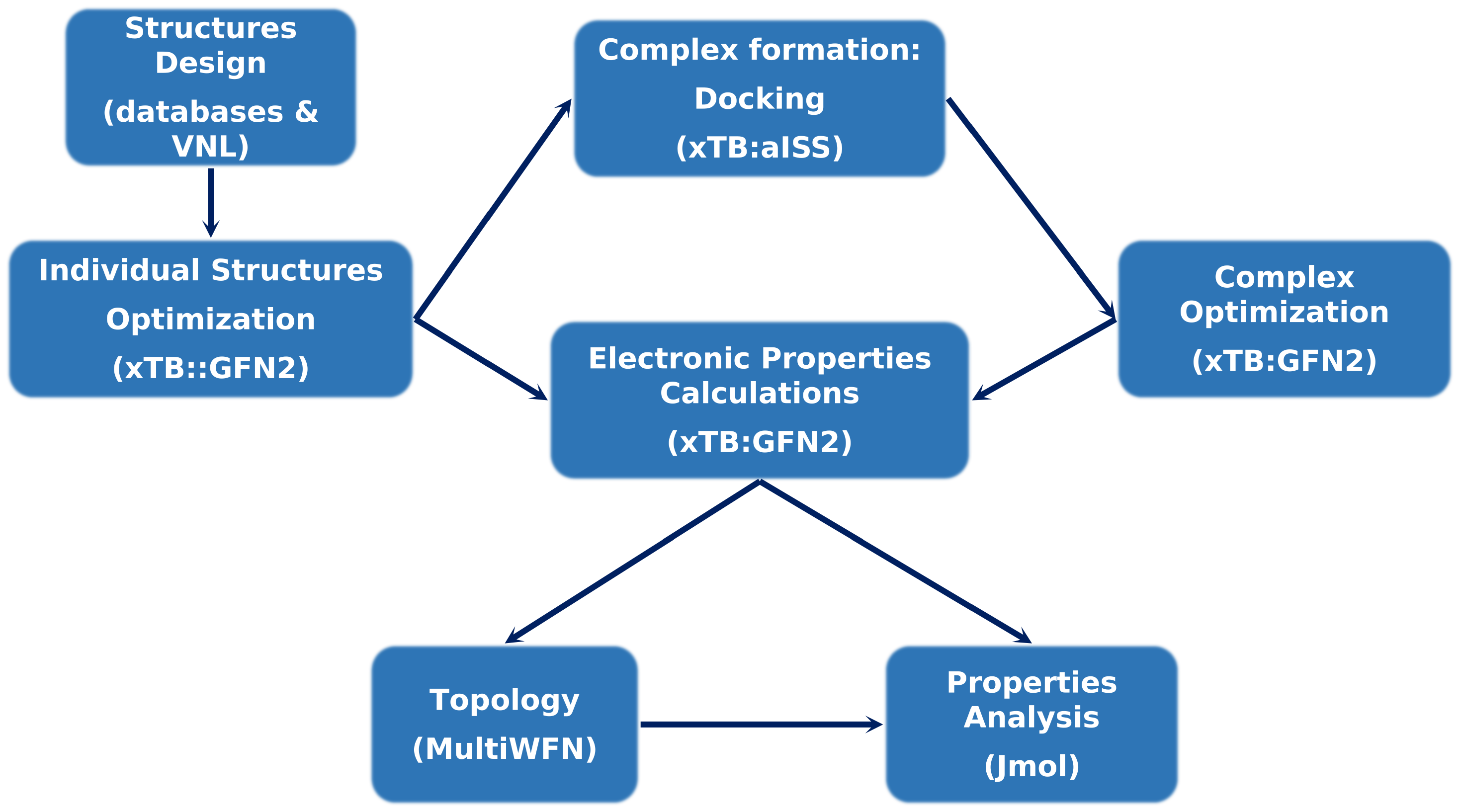}
\caption{Flowchart of the used methodology.}
\label{Fig:Method}
\end{figure}

The carbon nanobelt and Möbius nanobelt structures are abbreviated as CNB and MCNB, respectively. Complexes formed between these nanobelts and greenhouse gases are denoted as CNB+gas or MCNB+gas (e.g., CNB+NO). Figure~\ref{Fig:Method} illustrates the methodological workflow used in this study.

The calculations were performed using the xTB program, which implements a semiempirical tight-binding method. The xTB package's effectiveness has been thoroughly assessed across various databases encompassing transition metals, organometallics, and lanthanide complexes. When compared to more computationally intensive methods like coupled cluster and Density Functional Theory (DFT), the xTB package demonstrated remarkably accurate results \cite{xTB_2,xTB_GFN0,xTB_GFN2,xTB_1}.

The calculations followed the procedures detailed in the methodology flow chart (see Figure~\ref{Fig:Method}). First, the structures of each system, comprising two nanobelts and nine gases, were optimized. Next, the docking process was performed using automated Interaction Site Screening (aISS)~\cite{xTB-dock}. After identifying pockets in molecule A (the nanobelts), a three-dimensional (3D) screening was conducted to detect $\pi$-$\pi$ stacking interactions in various orientations. Subsequently, the global orientations of molecule B (the gases) were determined within an angular grid surrounding molecule A. Interaction energy (xTB-IFF) was then used to categorize the generated structures~\cite{xTB-IFF}. By default, one hundred structures with lower interaction energies were selected for a further two-step optimization using a genetic algorithm, ensuring incorporation of conformations missed in the initial screening. During this two-step genetic optimization, pair positions of molecule B were randomly crossed around molecule A. Then, 50\% of the structures underwent random mutations in both position and angle. After ten iterations of this comprehensive search process, ten complexes with the lowest interaction energy were selected. The structure with the lowest interaction energy was finally used as the input for the complex optimization.

Geometry optimization was performed using the GFN2-xTB method, a self-consistent approach that includes multipole electrostatics and density-dependent dispersion effects~\cite{xTB_GFN2}. Stringent optimization criteria were applied, with convergence thresholds of $5\times10^{-8}$~E\textsubscript{h} for energy and $5\times10^{-5}$~E\textsubscript{h}/a\textsubscript{0} for the gradient norm (where a\textsubscript{0} is the Bohr radius).

Employing a spin polarization approach, we computed various electronic properties, including the system's total energy, the energies of the highest occupied molecular orbital (HOMO, $\varepsilon_H$) and lowest unoccupied molecular orbital (LUMO, $\varepsilon_L$), the HOMO-LUMO energy gap ($\Delta \varepsilon=\varepsilon_H - \varepsilon_L$), as well as electron transfer integrals~\cite{DIPRO}.

In the context of gas adsorption on surfaces, sensitivity quantifies the responsiveness of the adsorption process to parameter changes. Here, the sensitivity factor ($\Delta \Delta \varepsilon$) is derived from the alteration in the nanobelt's electronic gap following gas adsorption, expressed as:

\begin{equation}
\label{Eq:Conductivity}
\Delta \Delta \varepsilon = \left| \Delta \varepsilon - \Delta \varepsilon_0 \right| / \Delta \varepsilon_0,
\end{equation}
where $\Delta \varepsilon_0$ and $\Delta \varepsilon$ are the electronic gap of the isolated nanobelts and complexes systems, respectively.

The sensitivity of a sensor can be quantified by measuring changes in the material's conductivity ($\Delta \sigma$), which correspond to variations in the electric signal~\cite{AbdalkareemJasim-Inorg.Chem.Commun.-146-110158-2022,Goel-EngineeringReports-5--2023}. The conductivity changes can be determined using the following calculations:

\begin{equation}
\label{Eq:Conductivity}
\Delta \sigma = (\sigma - \sigma_0)/\sigma_0,
\end{equation}
where $\sigma_0$ and $\sigma$ represent the conductivities of isolated nanobelts and the complex system, respectively. According to Kittel~\cite{Kittel1996}, the conductivity of a semiconductor is directly related to both the intrinsic carrier concentration (primarily electrons) and their mobility. Consequently, $\sigma$ is also proportional to the electronic gap:

\begin{equation}
\label{Eq:ConductivityGap}
\sigma \propto Exp(-\Delta \varepsilon/2k_BT),
\end{equation}
$k_B$ and $T$ are the Boltzmann constant and system temperature, respectively. Higher values of $\Delta \sigma$ indicate a higher sensitivity of the nanobelt to the corresponding gas.

To investigate the charge carrier mobility between nanobelts and adsorbed gas molecules, we calculated the electron transfer integrals using the dimer projection (DIPRO) method~\cite{xTB-DIPRO}. In our analysis, $J_{oc}$ denotes hole transport (occupied molecular orbitals), $J_{un}$ represents electron transport (unoccupied molecular orbitals), and $J$ signifies the total charge transfer, encompassing both hole and electron transport between occupied and unoccupied molecular orbitals, respectively. Larger $J$ values indicate stronger coupling between the two fragments, suggesting a more pronounced interaction between the nanobelt and gas molecule.

The adsorption energy ($E_{ads}$) of greenhouse gases on the carbon nanobelts was determined by utilizing the total system energies and applying the subsequent equation:

\begin{equation}
\label{Eq:bind}
E_{ads} = E_{NB+gas} - E_{NB}- E_{gas}.
\end{equation}

In equation~\ref{Eq:bind}, $E_{NB}$ and $E_{gas}$ are the energies
for the isolated nanobelts and the gas molecule, respectively, and
$E_{NB+gas}$ is the energy of the NB+gas complex (CNB+gas and MCNB+gas
systems).

Strong interactions, while enhancing adsorption, can inhibit gas detection by impeding desorption and prolonging recovery times. Recovery time, a crucial parameter for gas-sensing materials, exhibits an exponential relationship with adsorption energy, as predicted by transition theory:

\begin{equation}
\label{Eq:RecoveryTime}
\tau = \nu^{-1}_0 Exp (E_{ads}/k_BT)
\end{equation}
where $\nu_0$ is the exposed used frequency. In our investigation, the recovery time was calculated employing transition state theory, with parameters set to $\nu_0$=10\textsuperscript{12}~Hz (corresponding to ultraviolet radiation) and T=298.15~K.

The topological properties and descriptors, including critical points, electronic density, and Laplacian of the electronic density, were analyzed using MULTIWFN software~\cite{multiwfn}. This analysis was based on the wave function generated during the electronic property calculations.

Geometry optimization is a widely recognized technique that utilizes algorithms to locate a local minimum energy structure on the potential energy surface (PES). This process enables the identification of the most stable, low-energy conformations of a molecular system. However, it does not offer insights into the system's stability over time. In contrast, molecular dynamics (MD) simulations examine the motion of atoms and molecules at a specified temperature, such as 298.15 K, and provide a means to explore the PES, offering a more comprehensive understanding of the system's behavior.

We conducted molecular dynamics simulations under three distinct scenarios. The first scenario involved simulating each complex, using the structures obtained from the aISS step as initial conformations. This provided insight into the interactions between a single gas molecule and each nanobelt. In the second scenario, we utilized the PACKMOL software~\cite{packmol_0,packmol_1} to create initial complexes by adding 500 gas molecules to each nanobelt within a 20~\AA~radii spherical distribution, as illustrated in Figure~\ref{Fig:ComplexPackage}. This calculation allowed us to observe the evolution of a more complex system over time. The third scenario involved using PACKMOL to create a complex comprising 500 dry air molecules (consisting of 78\%~nitrogen, 21\%~oxygen, and 1\%~argon~\cite{Air}) and 25~greenhouse gas molecules within a 20~\AA~radii spherical distribution centered on the nanobelt, as depicted in Figure~\ref{Fig:ComplexPackageAir}. This calculation aimed to replicate a more realistic and complex system.

The molecular dynamics simulations were conducted with a production run time of 100~ps, utilizing a time step of 2~fs and a dump step of 50~fs, where the final configuration was recorded in a trajectory file. These calculations employed the GFN-FF force field, which is specifically designed to balance high computational efficiency with the accuracy typically associated with quantum mechanics methods~\cite{xTB_GFN-FF}.

\newcommand{\sizeA}{7.0cm}
\begin{figure}[htpb]
\centering
\subfigure[]{\includegraphics[height=\sizeA]{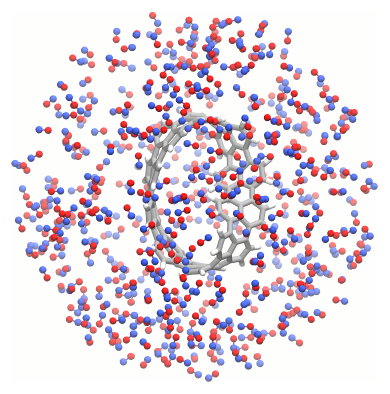}\label{Fig:ComplexPackage}}
\subfigure[]{\includegraphics[height=\sizeA]{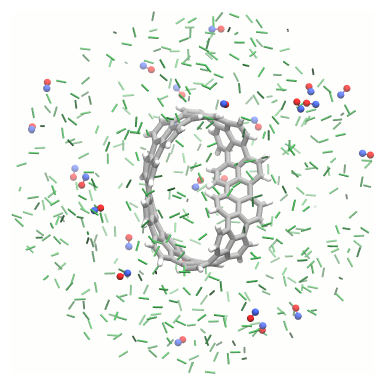}\label{Fig:ComplexPackageAir}}
\caption{\label{Fig:MD_ComplexPackages} (a) A MCNB packed with 500~NO molecules in a spherical conformation of 20~\AA~radii. (b) A MCNB packed with 500 dry air molecules (represented as green tubes) and with 25~NO molecules in a spherical conformation of 20~\AA~radii. Images rendered with VMD software~\cite{vmd,vmd_Tachyon} using the CPK color scheme.}
\end{figure}

Characterizing the particle distribution in heterogeneous systems can be achieved using the radial distribution function (RDF), denoted as $g(r)$. This pair correlation function provides insight into the average radial arrangement of particles around each other within the system~\cite{Hansen2006}. Offering a quantitative description of the spatial relationships between particles, the RDF can be calculated using the following expression~\ref{Eq:RDF}:

\begin{equation}
\label{Eq:RDF}
g(\bf{r}) = \frac{n(\bf{r})}{\rho 4 \pi \bf{r}^2 \Delta \bf{r}},
\end{equation}
where $n(\bf{r})$ is the mean number of particles in a shell of width $\Delta \bf{r}$ at distance $\bf{r}$, and $\rho$ is the mean particle density.

\section{RESULTS AND DISCUSSION}
\label{Sec:results}
\subsection{Gas adsorption at carbon nanobelts}
\label{Sec:Geometry}

Tables~\ref{Tab:ResultsAdsCNB} and~\ref{Tab:ResultsAdsMCNB} present the calculated parameters for the optimized geometry complexes of CNB and MCNB, respectively, with adsorption energies, $E_{ads}$, arranged in ascending order (more negative values indicating stronger adsorption). All studied gases exhibited negative adsorption energies, demonstrating favorable interactions with both nanobelt types. The gases NO, COCl\textsubscript{2}, and NO\textsubscript{2} showed the strongest interactions with both CNB and MCNB systems. Notably, the Möbius carbon-type nanobelt (MCNB) consistently demonstrated stronger adsorption energies for all gases.

The sensitivity factor ($\Delta \Delta \varepsilon$), presented in Tables~\ref{Tab:ResultsAdsCNB} and~\ref{Tab:ResultsAdsMCNB}, serves as an indicator of the nanobelts' response to gas adsorption, with larger values indicating enhanced sensitivity. Due to the experimental challenges in direct electronic gap measurements, we focused our analysis on material conductivity variations. The change in conductivity ($\Delta \sigma$) provides a practical measure of material sensitivity during molecular interactions, as it is intrinsically linked to the electronic gap ($\Delta \varepsilon$) according to Equation (\ref{Eq:ConductivityGap}). Using the conductivity of isolated nanobelts ($\sigma_0$) as a reference, Tables~\ref{Tab:ResultsAdsCNB} and~\ref{Tab:ResultsAdsMCNB} present both $\Delta \varepsilon$ and $\Delta \sigma$ values, where positive and negative $\Delta \sigma$ values represent conductivity enhancement and reduction, respectively. Both descriptors - the sensitivity factor ($\Delta \Delta \varepsilon$) and the conductivity variation ($\% \Delta \sigma$) - exhibit minimal differences across all adsorbed gases, suggesting negligible electrical response from both CNB and MCNB upon gas adsorption. This limited sensitivity can be attributed to the low molecular hardness ($\eta$) of CNB and MCNB~\cite{Camps_DiscoverNano2024}, as lower $\eta$ values facilitate electron redistribution throughout the molecular structure.

\begin{table}[htpb]
\caption{Adsorption energy ($E_{ads}$), HOMO ($\varepsilon_H$), LUMO
($\varepsilon_L$), gap ($\Delta \varepsilon$), sensitivity factor ($\Delta \Delta \varepsilon$), electrical conductivity variation ($\% \Delta \sigma$) by the adsorption, recovery time ($\tau$), and effective electron transfer integral ($J_{oc}$/$J_{un}$/$J$) for the CNB systems$^\dagger$.}
\label{Tab:ResultsAdsCNB}
\begin{center}
\setlength\extrarowheight{-3pt}
\begin{tabular}{lrrrrrrrrr}
  \hline
  System & $E_{ads}$ & $\varepsilon_H$ &
  $\varepsilon_L$ & $\Delta \varepsilon$ & $\Delta \Delta \varepsilon$ & $\% \Delta \sigma$  & $\tau$ &   $J_{oc}$/$J_{un}$/$J$ \\
  \hline
  \hline \\
  CNB                       & ---    & -8.936 & -0.312 & 8.624 & --- & --- & ---           & ---\\
  CNB+NO                    & -0.636 & -8.823 & -0.307 & 8.516 & 1   & 10  & 88.45~ms & ---   \\
  CNB+COCl\textsubscript{2} & -0.449 & -8.956 & -0.313 & 8.643 & 0   & -2  & 52.28~$\mu$s & 0/0/27    \\
  CNB+NO\textsubscript{2}   & -0.438 & -8.827 & -0.313 & 8.513 & 1   & 10  & 34.63~$\mu$s & 2/18/73   \\
  CNB+NH\textsubscript{3}   & -0.408 & -8.944 & -0.313 & 8.631 & 0   & -1  & 10.45~$\mu$s & 2/6/3     \\
  CNB+CO                    & -0.240 & -8.949 & -0.312 & 8.636 & 0   & -1  & 13.56~ns & 54/55/106 \\
  CNB+CO\textsubscript{2}   & -0.221 & -8.950 & -0.312 & 8.638 & 0   & -1  & 6.42~ns & 24/15/44  \\
  CNB+CH\textsubscript{3}OH & -0.213 & -8.950 & -0.312 & 8.638 & 0   & -1  & 4.67~ns & 3/6/40  \\
  CNB+H\textsubscript{2}S   & -0.210 & -8.960 & -0.313 & 8.647 & 0   & -2  & 4.03~ns & 9/1/3     \\
  CNB+CH\textsubscript{4}   & -0.126 & -8.935 & -0.312 & 8.623 & 0   &  0  & 0.14~ns & 22/4/10 \\
  \hline
\end{tabular}
\begin{flushleft}
\tiny {$^\dagger$ $E_{ads}$, $\varepsilon_H$,
$\varepsilon_L$, and $\Delta \varepsilon$, are in units of $eV$; $\Delta \Delta \varepsilon$ and $\% \Delta \sigma$ are dimensionless; and $J_{oc}$, $J_{un}$, and $J$ are in units of $meV$.}
\end{flushleft}
\end{center}
\end{table}

For practical sensor applications requiring re-usability, excessively strong adsorbate-surface interactions are undesirable as they impede efficient desorption. The sensor's re-usability is primarily determined by its recovery time -the duration needed for complete adsorbate removal. This recovery time ($\tau$) exhibits a direct relationship with the adsorption energy ($E_{ads}$) and can be quantitatively determined using Equation (\ref{Eq:RecoveryTime}). Lower $\tau$ values indicate more facile desorption of gas molecules from the nanobelt surface. Notably, the CNB system, despite having the highest $E_{ads}$, demonstrates the shortest recovery times, spanning from miliseconds to nanoseconds. Conversely, the MCNB system, characterized by lower $E_{ads}$ values, exhibits longer recovery times more than twice higher han for CNB system. Nevertheless, all observed recovery times remain within practically feasible ranges. The exception is the complex MCNB+NO, where te recovery times are of the order of hours.

For subsequent analyses, we define the best-ranked complexes as the three configurations with the lowest adsorption energies for each nanobelt type (CNB and MCNB).

\begin{table}[htpb]
\caption{Adsorption energy ($E_{ads}$), HOMO ($\varepsilon_H$), LUMO
($\varepsilon_L$), gap ($\Delta \varepsilon$), sensitivity factor ($\Delta \Delta \varepsilon$), electrical conductivity
variation ($\%\Delta \sigma$) by the adsorption, recovery time ($\tau$), and effective electron transfer integral ($J_{oc}$/$J_{un}$/$J$) for the MCNB systems$^\dagger$.}
\label{Tab:ResultsAdsMCNB}
\begin{center}
\setlength\extrarowheight{-3pt}
\begin{tabular}{lrrrrrrrr}
  \hline
  System & $E_{ads}$ & $\varepsilon_H$ &
  $\varepsilon_L$ & $\Delta \varepsilon$ & $\Delta \Delta \varepsilon$ & $\% \Delta \sigma$  & $\tau$ &
  $J_{oc}$/$J_{un}$/$J$ \\
  \hline
  \hline \\
  MCNB                       & ---    & -8.863 & -0.313 & 8.550 & ---& --- & ---         & ---\\
  MCNB+NO                    & -1.595 & -8.761 & -0.310 & 8.450 & 1 &  9 & ---           & ---  \\
  MCNB+COCl\textsubscript{2} & -0.669 & -8.881 & -0.313 & 8.568 & 0 & -1 & 321.73~ms     & 16/2/16  \\
  MCNB+NO\textsubscript{2}   & -0.637 & -8.810 & -0.314 & 8.496 & 1 &  5 & 91.85~ms      & 13/1/25  \\
  MCNB+NH\textsubscript{3}   & -0.519 & -8.861 & -0.313 & 8.549 & 0 &  0 & 855.19~$\mu$s & 15/16/30  \\
  MCNB+CO\textsubscript{2}   & -0.372 & -8.882 & -0.313 & 8.568 & 0 & -2 & 2.55~$\mu$s   & 29/13/34  \\
  MCNB+CH\textsubscript{3}OH & -0.363 & -8.866 & -0.313 & 8.553 & 0 &  0 & 1.75~$\mu$s   & 4/14/59  \\
  MCNB+H\textsubscript{2}S   & -0.349 & -8.873 & -0.313 & 8.560 & 0 & -1 & 1.02~$\mu$s   & 13/7/6  \\
  MCNB+CO                    & -0.331 & -8.873 & -0.313 & 8.560 & 0 & -1 & 0.50~$\mu$s   & 32/25/88  \\
  MCNB+CH\textsubscript{4}   & -0.223 & -8.860 & -0.313 & 8.548 & 0 &  0 & 6.81~ns       & 30/40/73  \\
  \hline
\end{tabular}
\begin{flushleft}
\tiny {$^\dagger$ $E_{ads}$, $\varepsilon_H$,
$\varepsilon_L$, and $\Delta \varepsilon$, are in units of $eV$; $\Delta \Delta \varepsilon$ and $\% \Delta \sigma$ are dimensionless; and $J_{oc}$, $J_{un}$, and $J$ are in units of $meV$.}
\end{flushleft}
\end{center}
\end{table}
\subsection{Electronic properties}
\label{Sec:ElecProp}

Figures~\ref{Fig:MO_CNB} and~\ref{Fig:MO_MCNB} illustrate the optimized structures, and the calculated frontier orbitals -HOMO (middle row) and LUMO (bottom row)- for both pristine nanobelts and their complexes with adsorbed gases.

The CNB displays a homogeneous distribution of molecular orbitals throughout its belt structure, a characteristic attributed to its symmetrical design. This distribution can be modified depending on the strength of interactions with adsorbed gases. Even for the strongest interaction case, i.e., CNB+NO, there is no significant redistribution of the belt wave functions except around the interaction region. In this case, the DIPRO analysis detected a single fragment indicating covalent bond formation between CNB and NO. This is further supported by the topological analysis presented in Section~\ref{Sec:Topo}.

The effective electron transfer integrals, that indicate the electron transfer between the gas and belt occupied/unoccupied orbitals, are not null (see Tables~\ref{Tab:ResultsAdsCNB} and~\ref{Tab:ResultsAdsMCNB}). Even with values different from zero, the interactions of COCl\textsubscript{2} and NO\textsubscript{2} with CNB, produced minor modifications to the orbital surface distribution. This can be related with the electronic stability of carbon-based nanobelts where the carbon atoms share a $sp^2$ hybridization with highly electron delocalization.

\renewcommand{\sizeA}{2.0cm}
\begin{figure}[tbph]
\centering
\begin{tabular}{cccc}
CNB & CNB+NO & CNB+COCl\textsubscript{2} & CNB+NO\textsubscript{2} \\

\subfigure[]{\includegraphics[height=\sizeA]{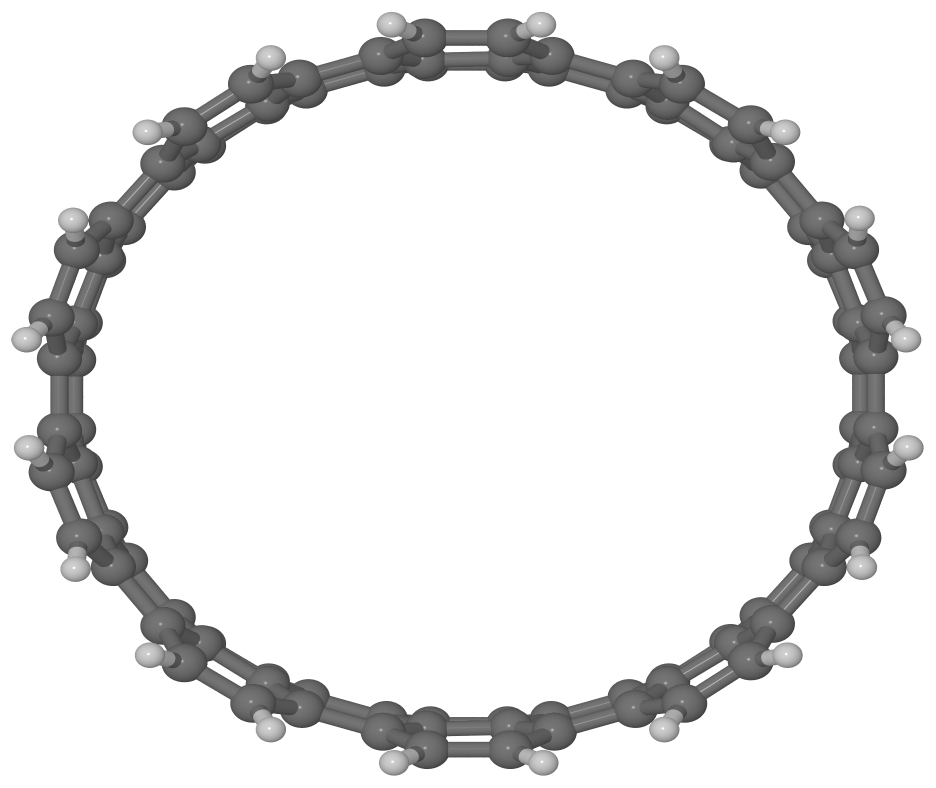}\label{Fig:CNB}}                         &
\subfigure[]{\includegraphics[height=\sizeA]{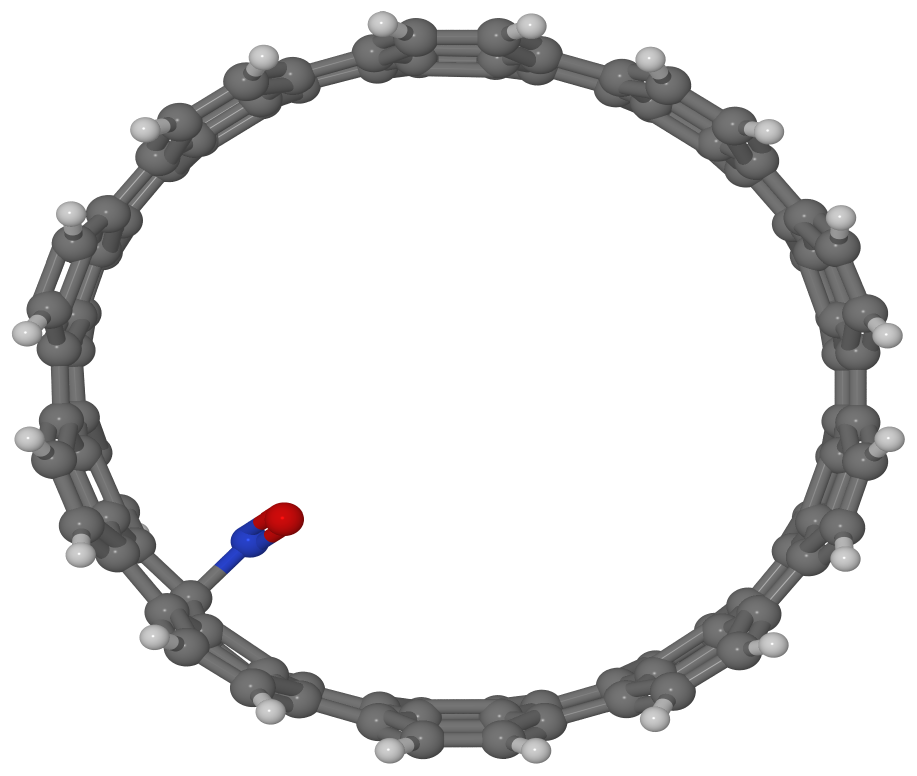}\label{Fig:CNB+NO}}                   &
\subfigure[]{\includegraphics[height=\sizeA]{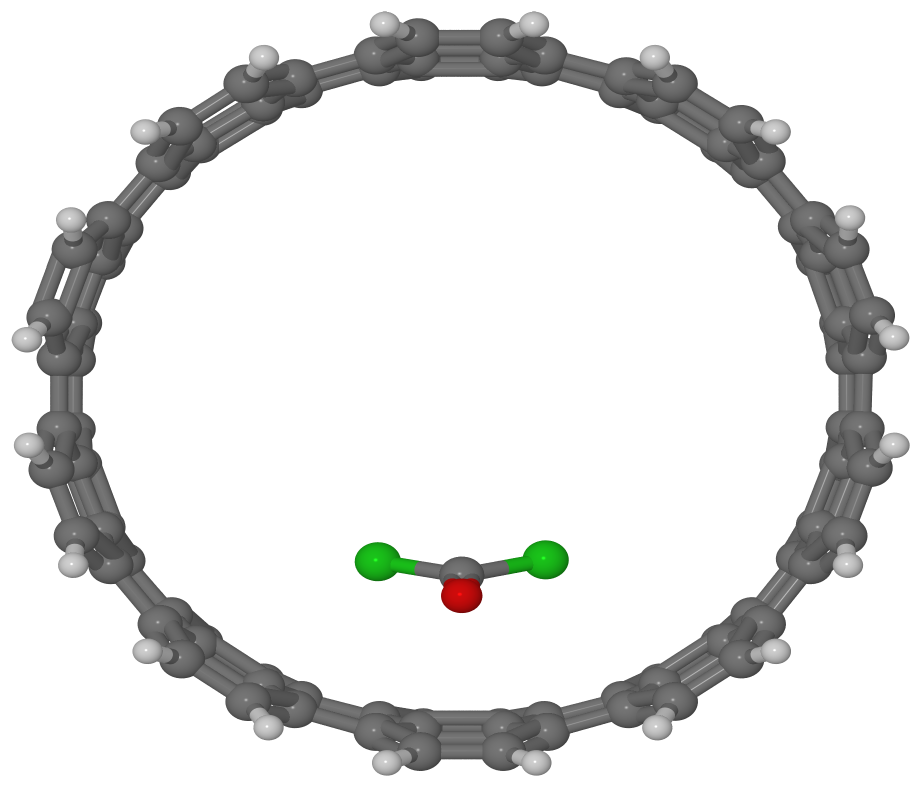}\label{Fig:CNB+COCl2}}             &
\subfigure[]{\includegraphics[height=\sizeA]{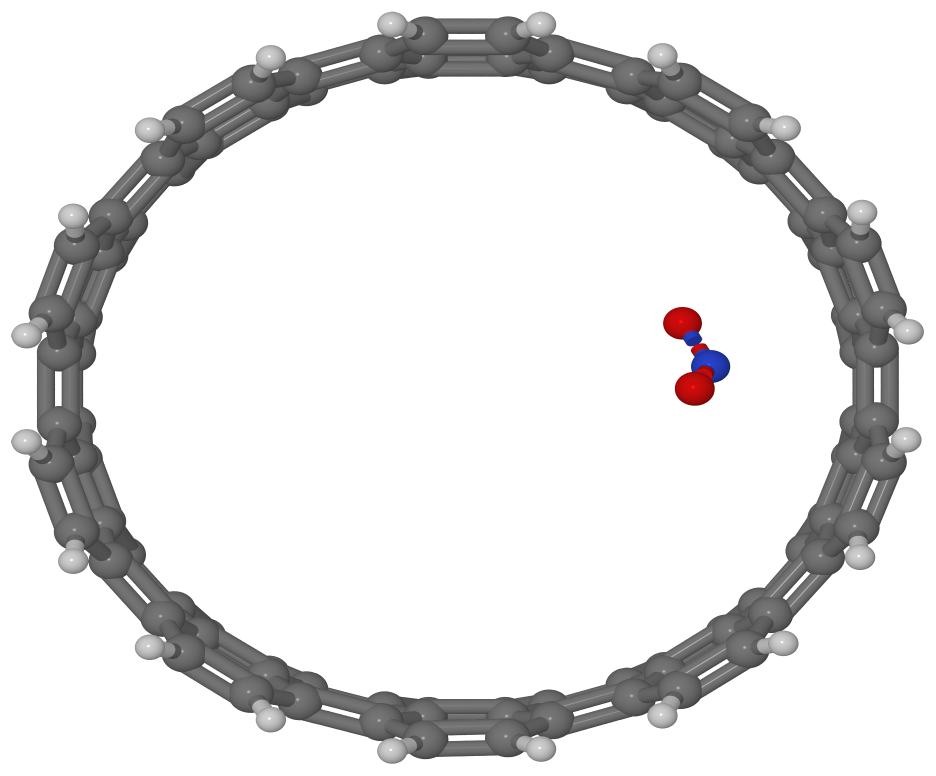}\label{Fig:CNB+NO2}}                \\
\subfigure[]{\includegraphics[height=\sizeA]{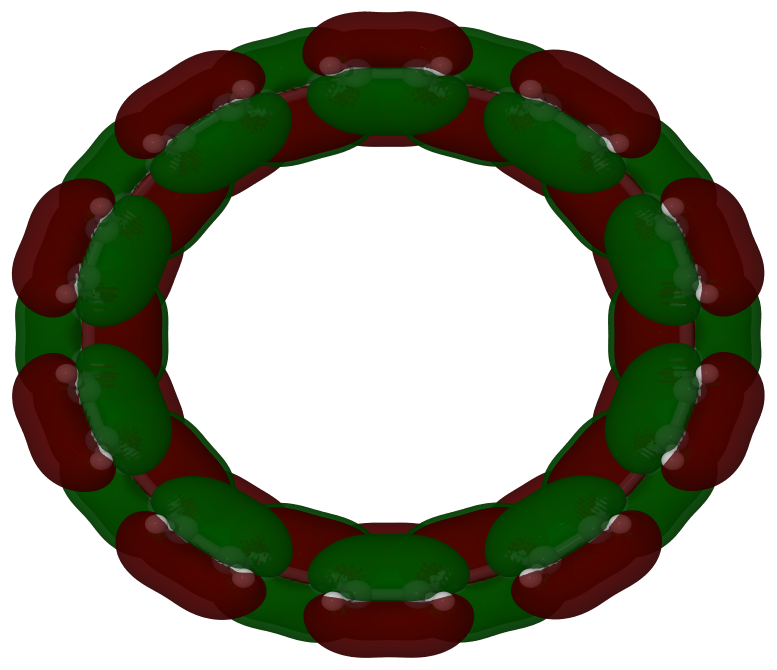}\label{Fig:HOMO_CNB}}               &
\subfigure[]{\includegraphics[height=\sizeA]{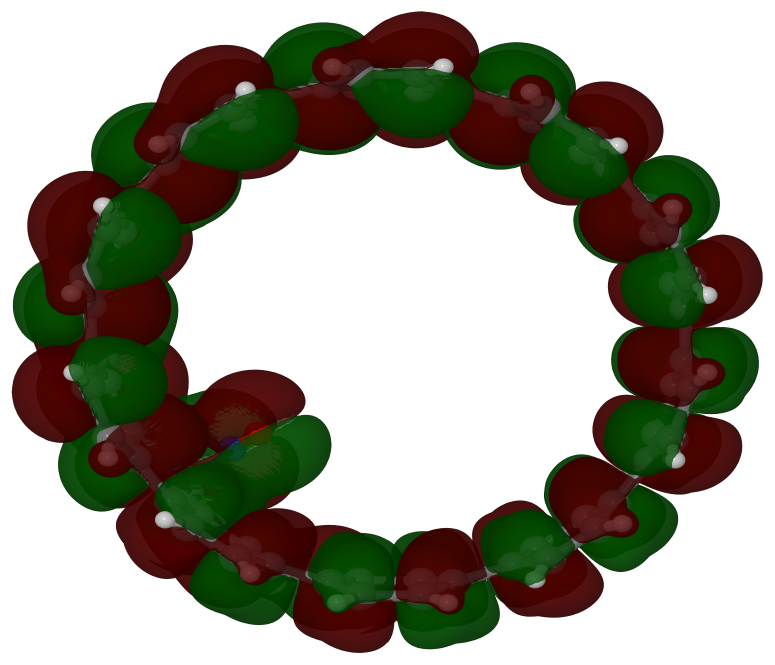}\label{Fig:HOMO_CNB+NO}}         &
\subfigure[]{\includegraphics[height=\sizeA]{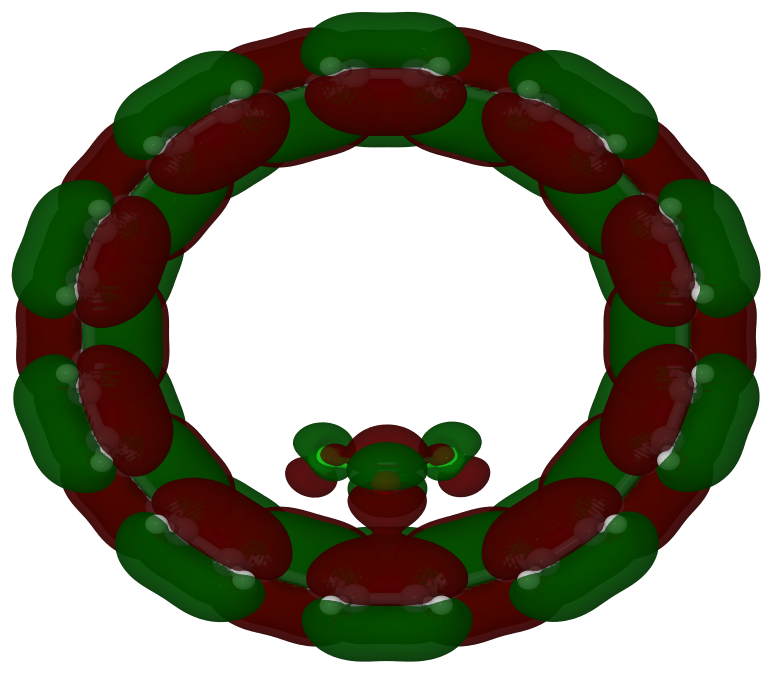}\label{Fig:HOMO_CNB+COCl2}}   &
\subfigure[]{\includegraphics[height=\sizeA]{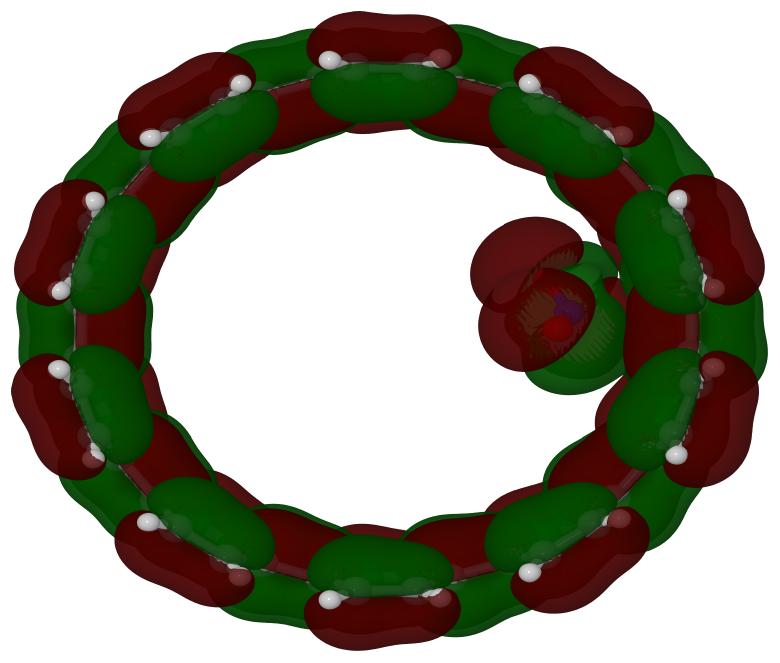}\label{Fig:HOMO_CNB+NO2}}       \\
\subfigure[]{\includegraphics[height=\sizeA]{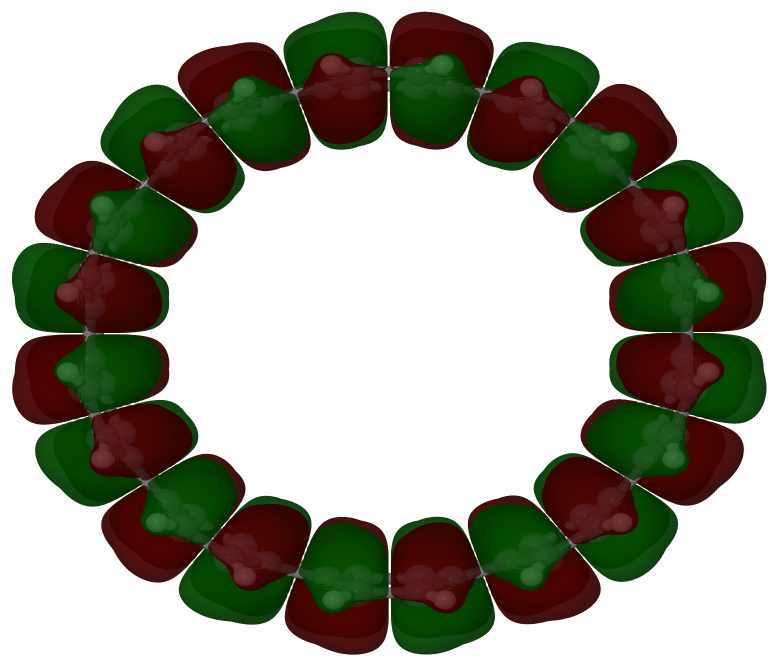}\label{Fig:LUMO_CNB}}                &
\subfigure[]{\includegraphics[height=\sizeA]{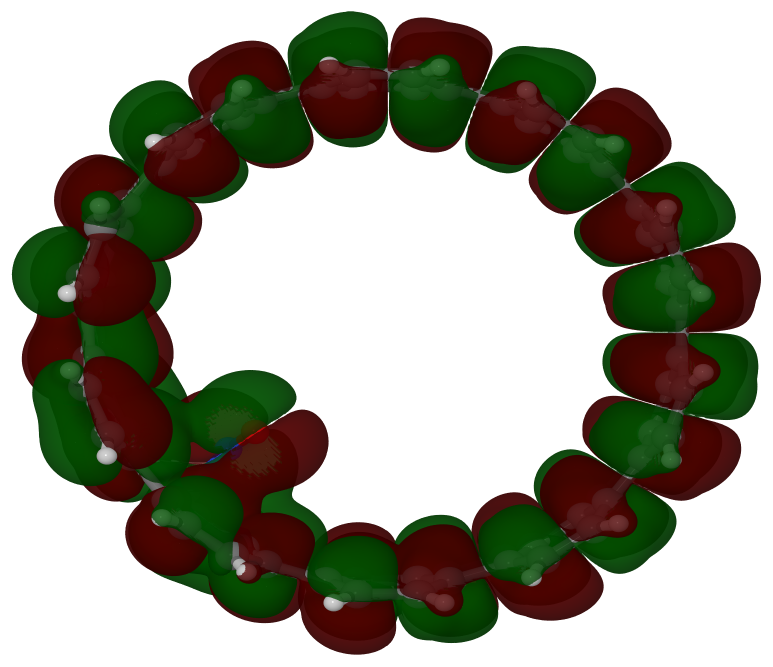}\label{Fig:LUMO_CNB+NO}}         &
\subfigure[]{\includegraphics[height=\sizeA]{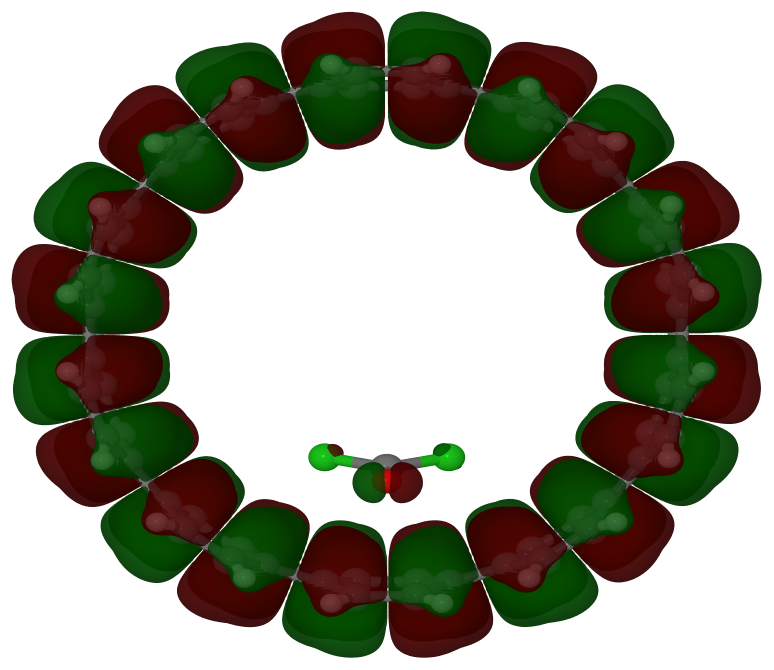}\label{Fig:LUMO_CNB+COCl2}}   &
\subfigure[]{\includegraphics[height=\sizeA]{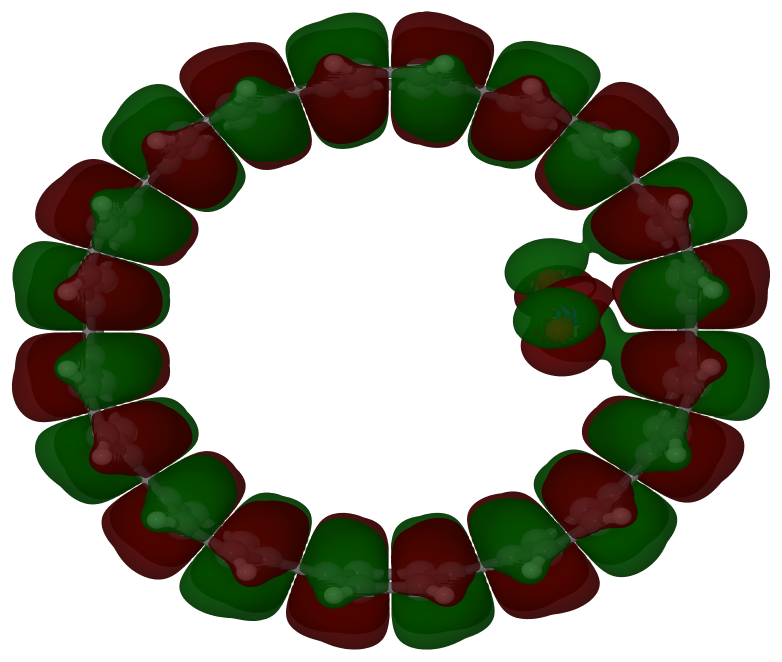}\label{Fig:LUMO_CNB+NO2}}      \\
\end{tabular}
\caption{\label{Fig:MO_CNB} Optimized structure (top row), HOMO (middle row) and LUMO (bottom row) for the CNB complexes and the most ranked. The red (green) represents negative (positive) values. Orbital surfaces were rendered with an isovalue equal to 0.001 and with Jmol software~\cite{jmol} using the CPK color scheme for atoms.}
\end{figure}

The inherent twist in the Möbius nanobelt significantly alter the frontier orbital distribution, with increased electron density concentrated around the twisted region. In systems like boron-nitride nanobelts~\cite{Camps_DiscoverNano2024,Camps_SurfInter2024}, these modifications are more pronounced on the HOMO surface compared to the LUMO surface. In our case, the best-ranked complexes show visible differences only on the region where the gas was absorbed. As mentioned previously, this can related with the low values of molecular hardness of carbon based nanobelts~\cite{Camps_DiscoverNano2024}.

\renewcommand{\sizeA}{2.50cm}
\begin{figure}[tbph]
\centering
\begin{tabular}{cccc}
MCNB & MCNB+NO & MCNB+COCl\textsubscript{2} & MCNB+NO\textsubscript{2} \\

\subfigure[]{\includegraphics[height=\sizeA]{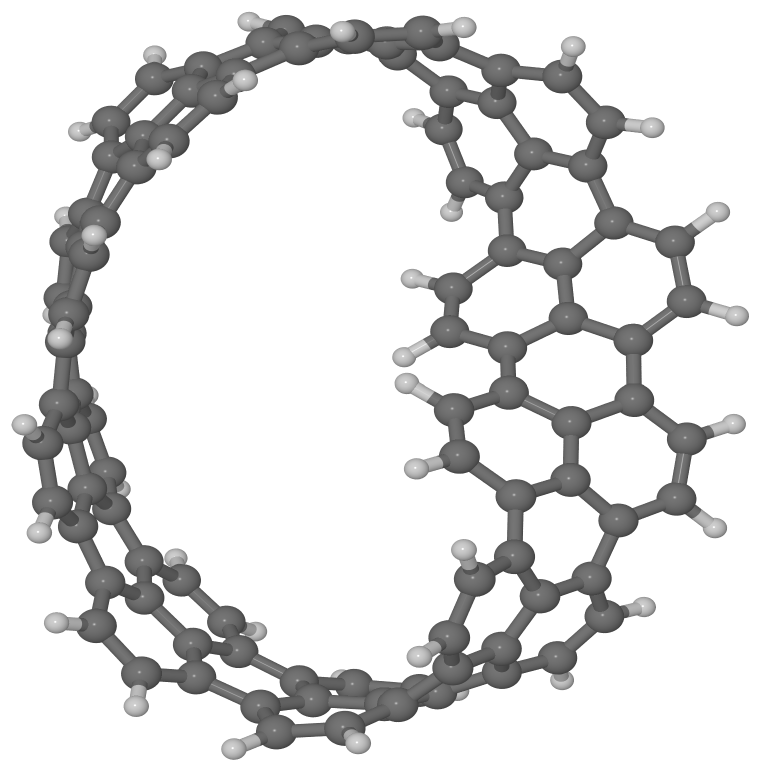}\label{Fig:MCNB}}                       &
\subfigure[]{\includegraphics[height=\sizeA]{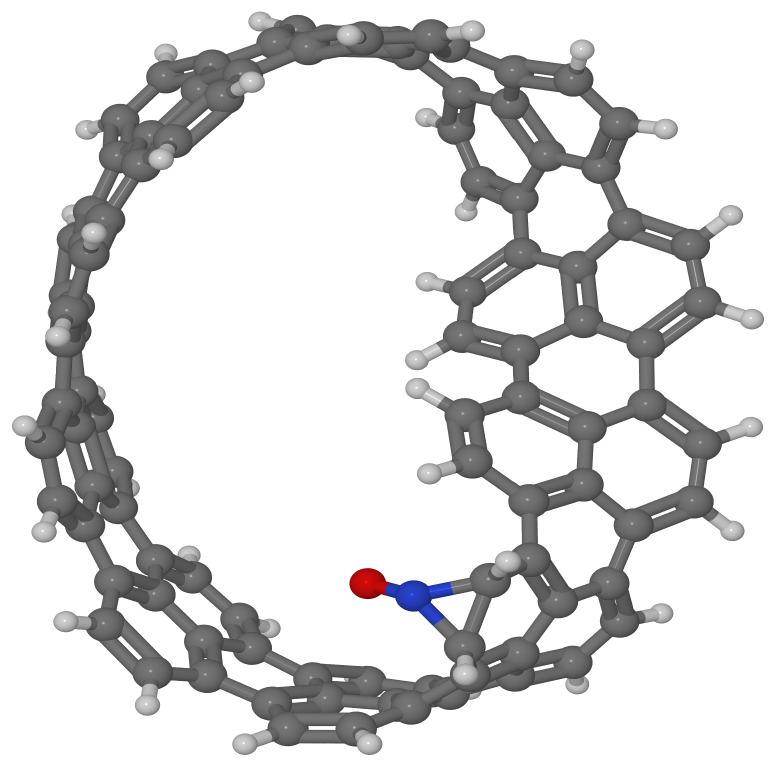}\label{Fig:MCNB+NO}}                 &
\subfigure[]{\includegraphics[height=\sizeA]{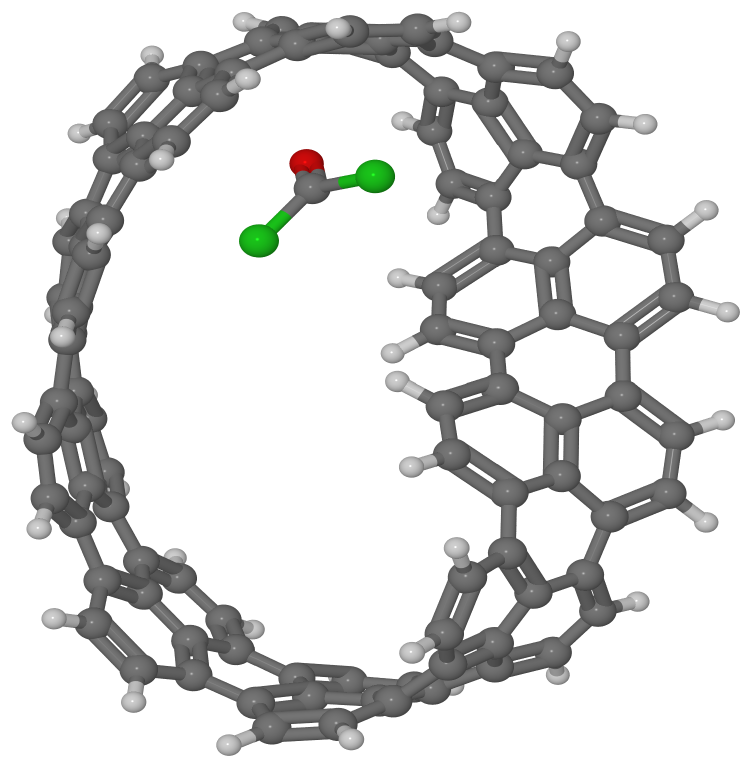}\label{Fig:MCNB+COCl2}}           &
\subfigure[]{\includegraphics[height=\sizeA]{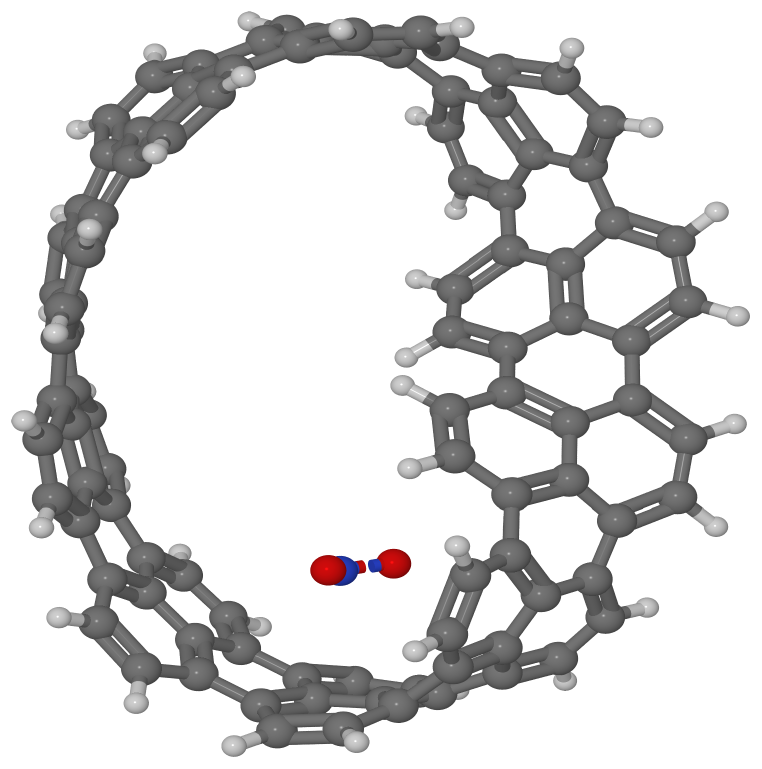}\label{Fig:MCNB+NO2}}               \\
\subfigure[]{\includegraphics[height=\sizeA]{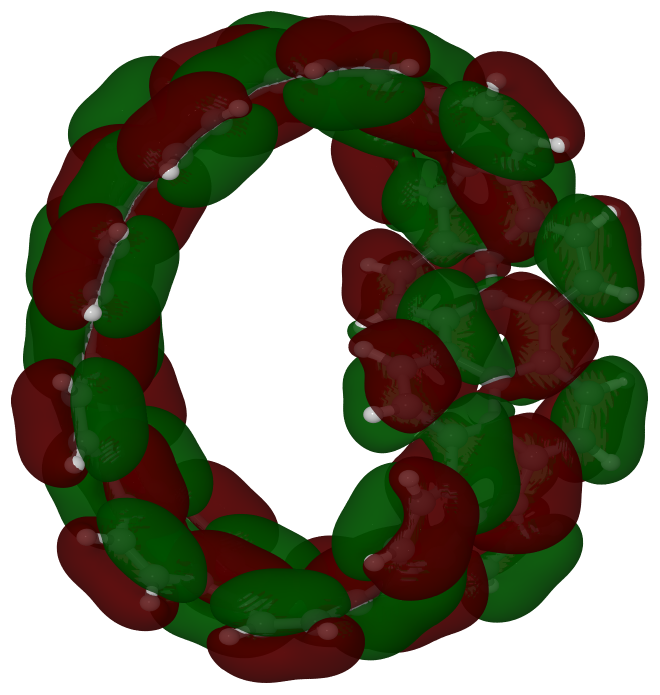}\label{Fig:HOMO_MCNB}}             &
\subfigure[]{\includegraphics[height=\sizeA]{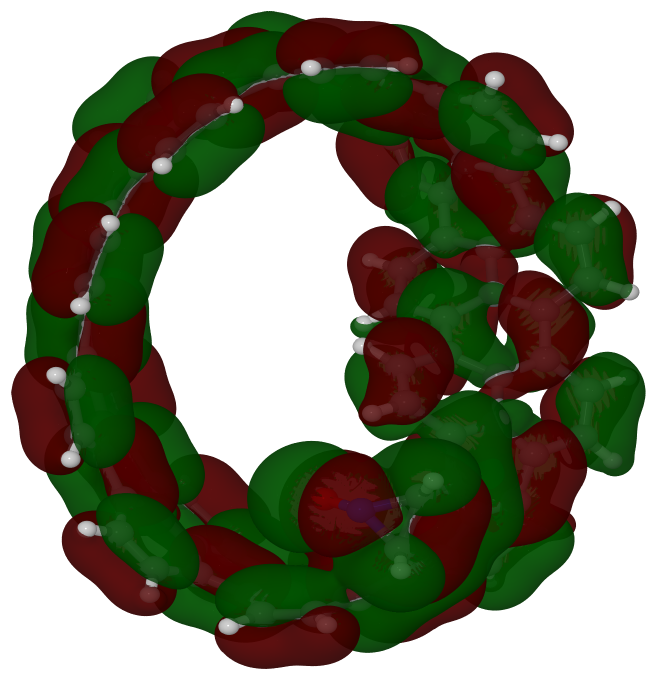}\label{Fig:HOMO_MCNB+NO}}       &
\subfigure[]{\includegraphics[height=\sizeA]{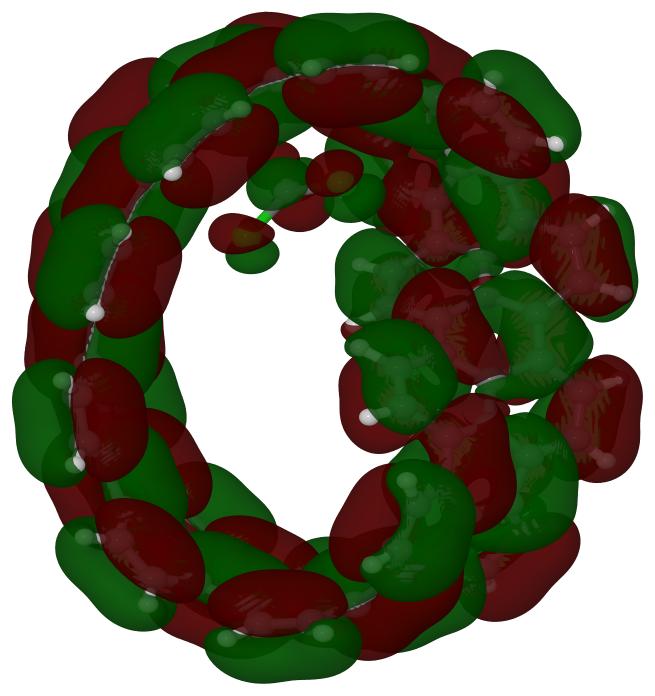}\label{Fig:HOMO_MCNB+COCl2}} &
\subfigure[]{\includegraphics[height=\sizeA]{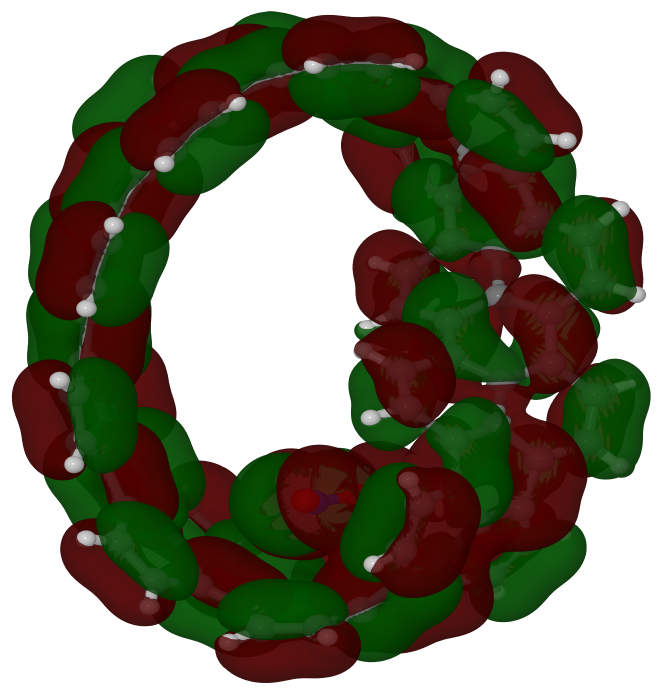}\label{Fig:HOMO_MCNB+NO2}}     \\
\subfigure[]{\includegraphics[height=\sizeA]{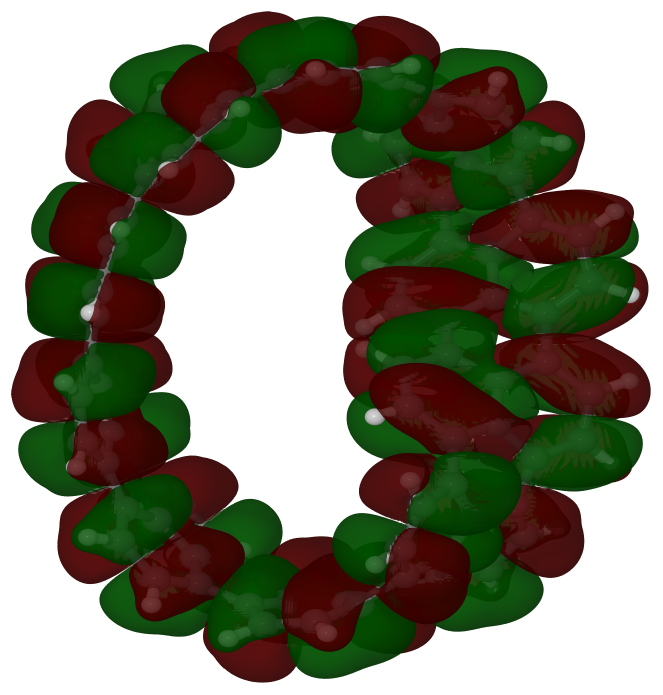}\label{Fig:LUMO_MCNB}}             &
\subfigure[]{\includegraphics[height=\sizeA]{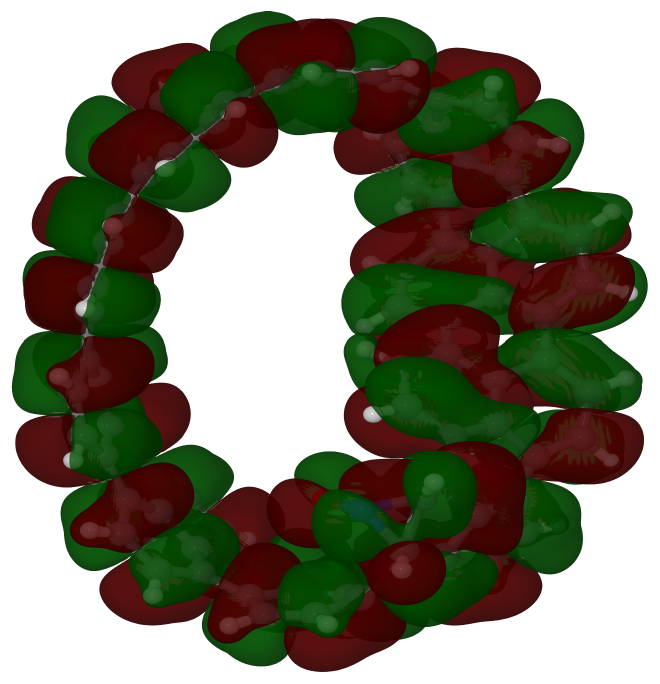}\label{Fig:LUMO_MCNB+NO}}       &
\subfigure[]{\includegraphics[height=\sizeA]{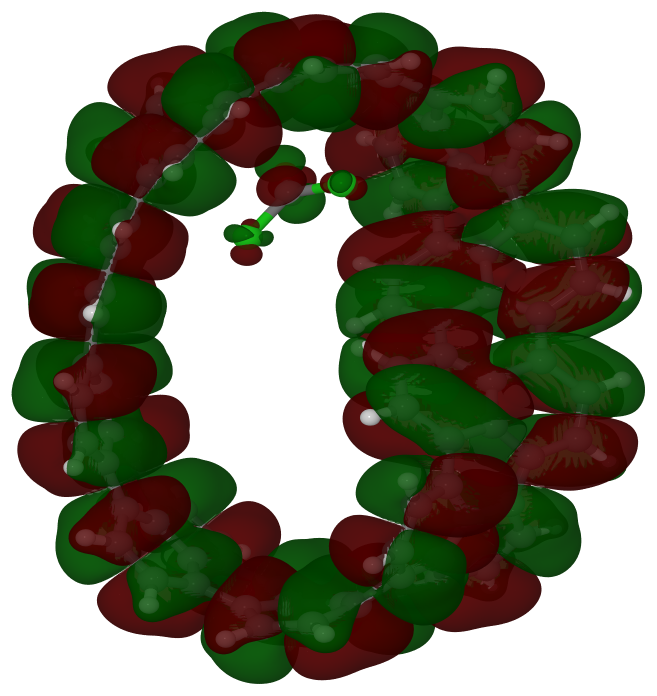}\label{Fig:LUMO_MCNB+COCl2}}  &
\subfigure[]{\includegraphics[height=\sizeA]{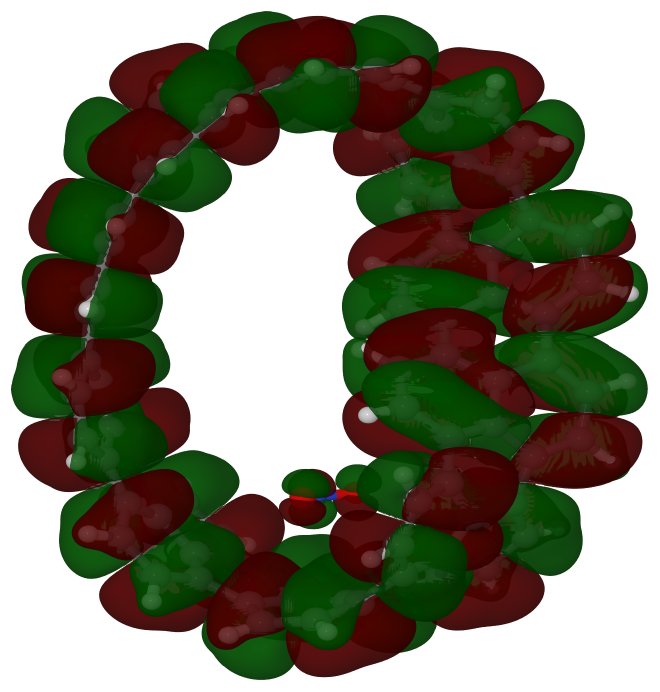}\label{Fig:LUMO_MCNB+NO2}}    \\
\end{tabular}
\caption{\label{Fig:MO_MCNB} Optimized structure (top row), HOMO (middle row) and LUMO (bottom row) for the MCNB complexes and the most ranked. The red (green) represents negative (positive) values. Orbital surfaces were rendered with an isovalue equal to 0.001 and with Jmol software~\cite{jmol} using the CPK color scheme for atoms.}
\end{figure}

\subsection{Topological analysis}
\label{Sec:Topo}

Topological analysis identifies critical points in the electron density field where the gradient norm vanishes. These points are categorized into four distinct types based on the number of negative eigenvalues of the Hessian matrix~\cite{Bader1994}. Details of the bond classification criteria are provided in Ref.~\cite{myMethods}. To characterize chemical bonding, we employ multiple quantum-mechanical descriptors: electron density ($\rho$), its Laplacian ($\nabla^2 \rho$), electron localization function (ELF), and localized orbital locator (LOL) at each bond critical point (BCP). The ELF, ranging from 0 to 1~\cite{elf,elf2}, quantifies electron pair localization, with higher values indicating stronger covalent character. Similarly, the LOL index (0-1) maps electron localization domains~\cite{lol}, typically showing smaller values at molecular boundaries and larger values in interior regions. Together, these metrics provide comprehensive insight into the nature of both covalent and non-covalent interactions in molecular systems.

From the topological calculations we obtain that the complexes CNB+NO, CNB+COCl\textsubscript{2}, and CNB+NO\textsubscript{2} have 1, 7, and 2 BCPs, and the MCNB+NO, MCNB+COCl\textsubscript{2}, and MCNB+NO\textsubscript{2} complexes have 3, 10, and 5 BCPs, respectively. The bond type (covalent or non-covalent) can be classified based on the electron density ($\rho$) and its Laplacian ($\nabla^2 \rho$) values. Bonds with $\rho > 0.20~a.u.$ and $\nabla^2 \rho < 0$ are characterized as covalent, while those with $\rho < 0.10~a.u.$ and $\nabla^2 \rho > 0$ indicate non-covalent interactions~\cite{Matta2007}.

Figure~\ref{Fig:RHO} shows the electron density, $\rho$, calculated for the best-ranked complexes which BCP has the highest value of $\rho$. In this figure, the light-blue line enclosing the systems represents the van der Waals surface; the thin black lines enclose different electronic density levels; the brown lines represent the bond path connecting two atoms and the blue dots represent the bonds critical point position. For each complexes, the bond path between both nanobelts and the NO, COCl\textsubscript{2}, and NO\textsubscript{2} gases, are shown.

\renewcommand{\sizeA}{3.0cm}
\begin{figure}[tbph]
\centering
\begin{tabular}{ccc}
\subfigure[CNB+NO]{\includegraphics[height=\sizeA]{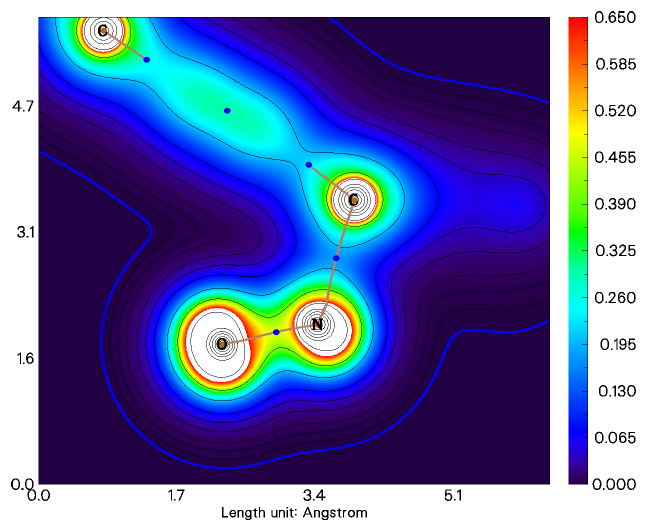}\label{Fig:RHO_CNB+NO}}                             &
\subfigure[CNB+COCl\textsubscript{2}]{\includegraphics[height=\sizeA]{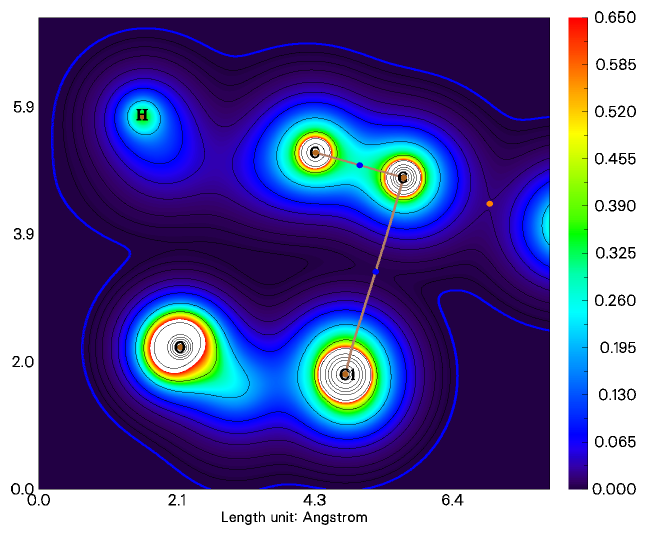}\label{Fig:RHO_CNB+COCl2}}    &
\subfigure[CNB+NO\textsubscript{2}]{\includegraphics[height=\sizeA]{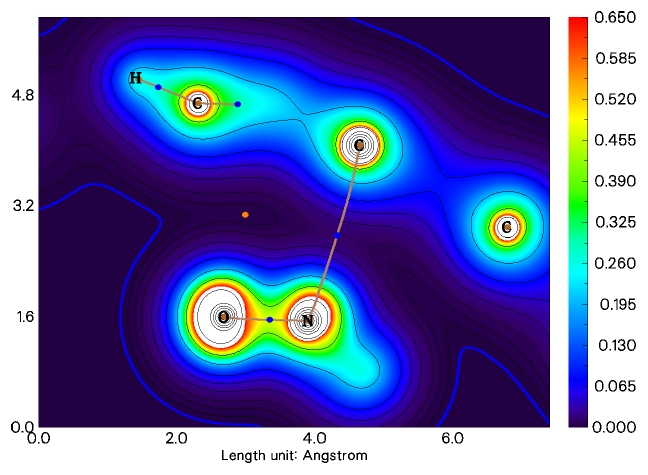}\label{Fig:RHO_CNB+NO2}}          \\
\subfigure[MCNB+NO]{\includegraphics[height=\sizeA]{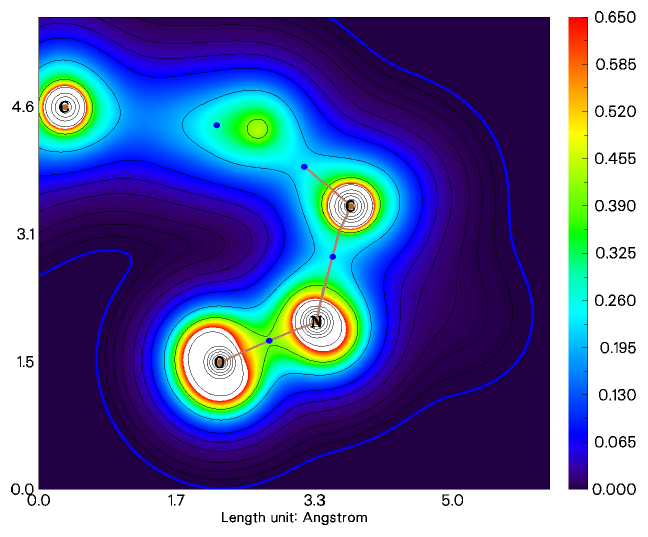}\label{Fig:RHO_MCNB+NO}}                          &
\subfigure[MCNB+COCl\textsubscript{2}]{\includegraphics[height=\sizeA]{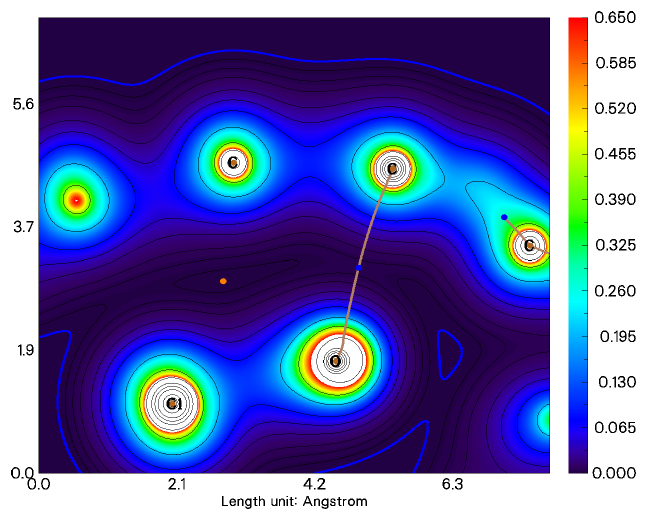}\label{Fig:RHO_MCNB+COCl2}} &
\subfigure[MCNB+NO\textsubscript{2}]{\includegraphics[height=\sizeA]{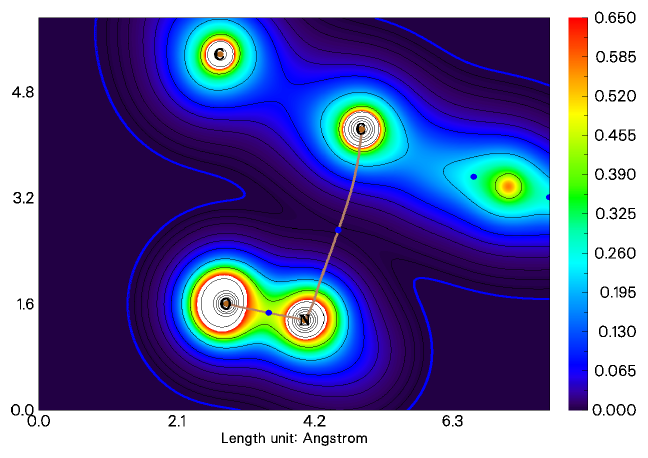}\label{Fig:RHO_MCNB+NO2}}       \\
\end{tabular}
\caption{\label{Fig:RHO} 2D representation of the electron density for the complexes with NO, COCl\textsubscript{2} and NO\textsubscript{2}.}
\end{figure}

\begin{figure}[htpb]
\centering
\includegraphics[width=14cm]{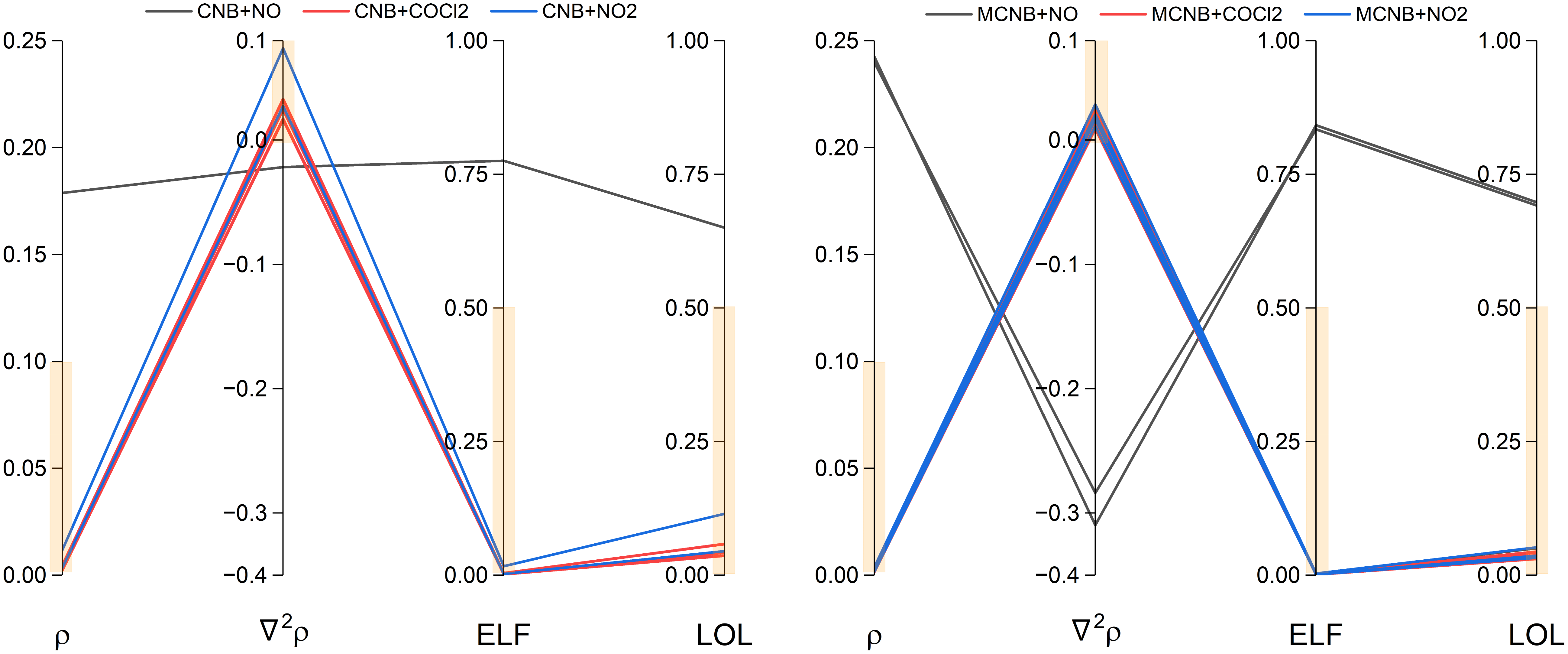}
\caption{Topological descriptors, $\rho$, $\nabla^2 \rho$, ELF and LOL, at the bond critical points (BCP) for each system.}
\label{Fig:TopoDesc}
\end{figure}

In order to characterize these bonds, the numerical values of the descriptors are shown in Figure~\ref{Fig:TopoDesc}. The black line represent the BCPs for complexes with NO, the red line, for complexes with COCl\textsubscript{2}, and the blue line, for complexes with NO\textsubscript{2}. Accordingly to previous classification (for the electron density ($\rho$) and its Laplacian ($\nabla^2 \rho$), the non-covalent regions are marked in orange for each descriptor. In case of ELF and LOL, we marked the regions below 0.5.

The CNB made one covalent bond with NO and all others bonds can be classified as non-covalent. In the case of MCNB, it made two covalent bonds with NO and all the others bonds are non-covalent. Comparing the values of all the descriptors, we can confirm that the MCNB interact strongly with the gases than the CNB.

\subsection{Molecular dynamics simulation}
\label{Sec:MD}
Molecular dynamics (MD) simulations were performed on the energetically optimized geometries of all complexes. The results revealed distinct binding behaviors: in the case of carbon nanobelts (CNBs), four gas molecules - NO, COCl\textsubscript{2}, NO\textsubscript{2}, and CO - maintained stable binding interactions with the nanobelt structure, while other gas molecules dissociated from the complex. For Möbius carbon nanobelts (MCNBs), enhanced stability was observed, with only MCNB-NH\textsubscript{3} and MCNB-CH\textsubscript{4} complexes exhibiting dissociation. These findings suggest that MCNBs demonstrate superior interaction capabilities with the investigated greenhouse gases.

\begin{figure}[htpb]
\centering
\includegraphics[width=14cm]{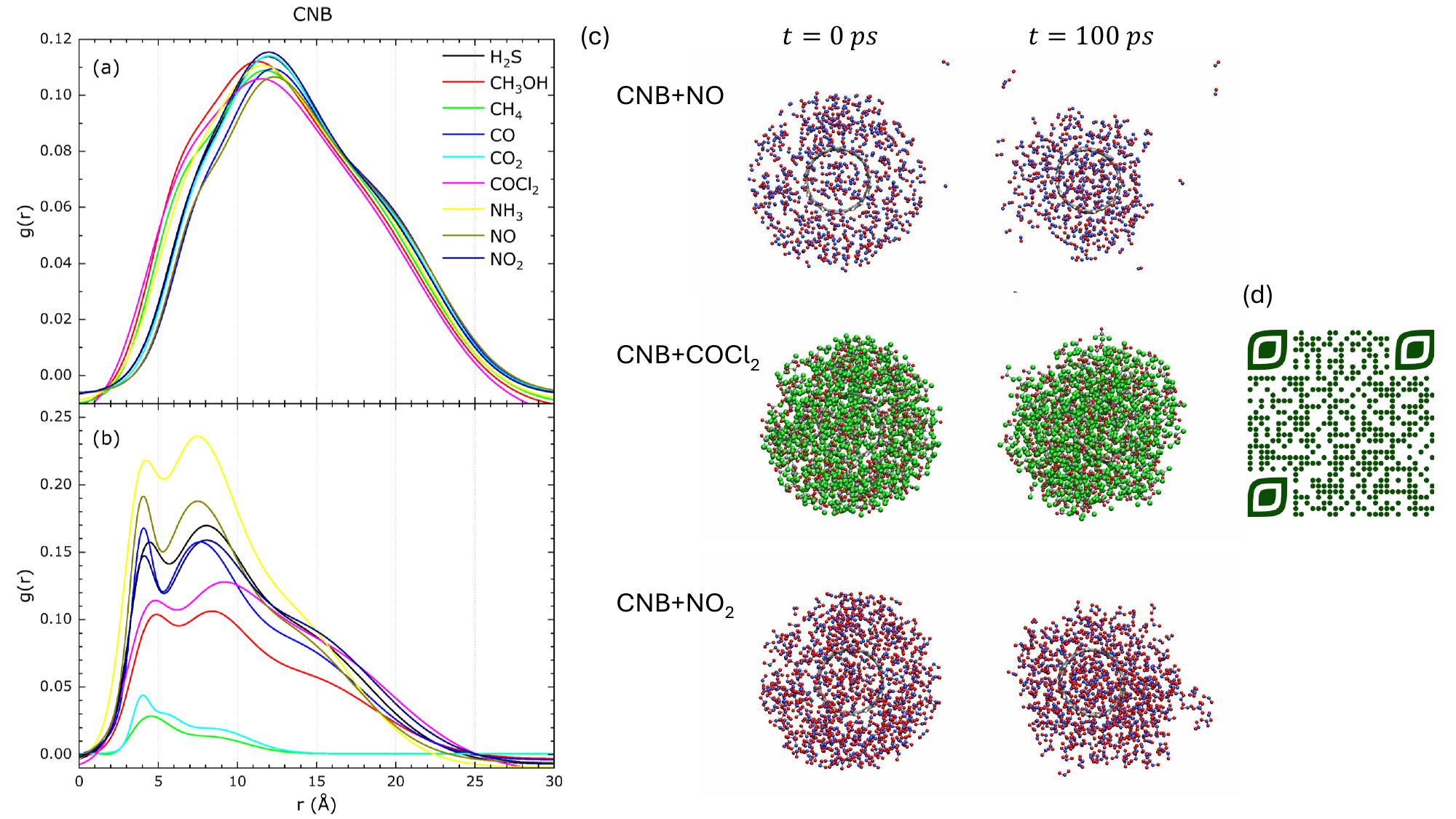}
\caption{(a) Initial, and (b), final radial distribution functions for all CNB complexes calculated with VMD~\cite{vmd,topotools}. (c) Initial (t=0~ps) and final (t=100~ps) configurations for the best--ranked complexes. (d) QR~Code: molecular dynamics animations for nanobelts interacting with~500 gas molecules.}
\label{Fig:RDF_CNB}
\end{figure}

\begin{figure}[htpb]
\centering
\includegraphics[width=14cm]{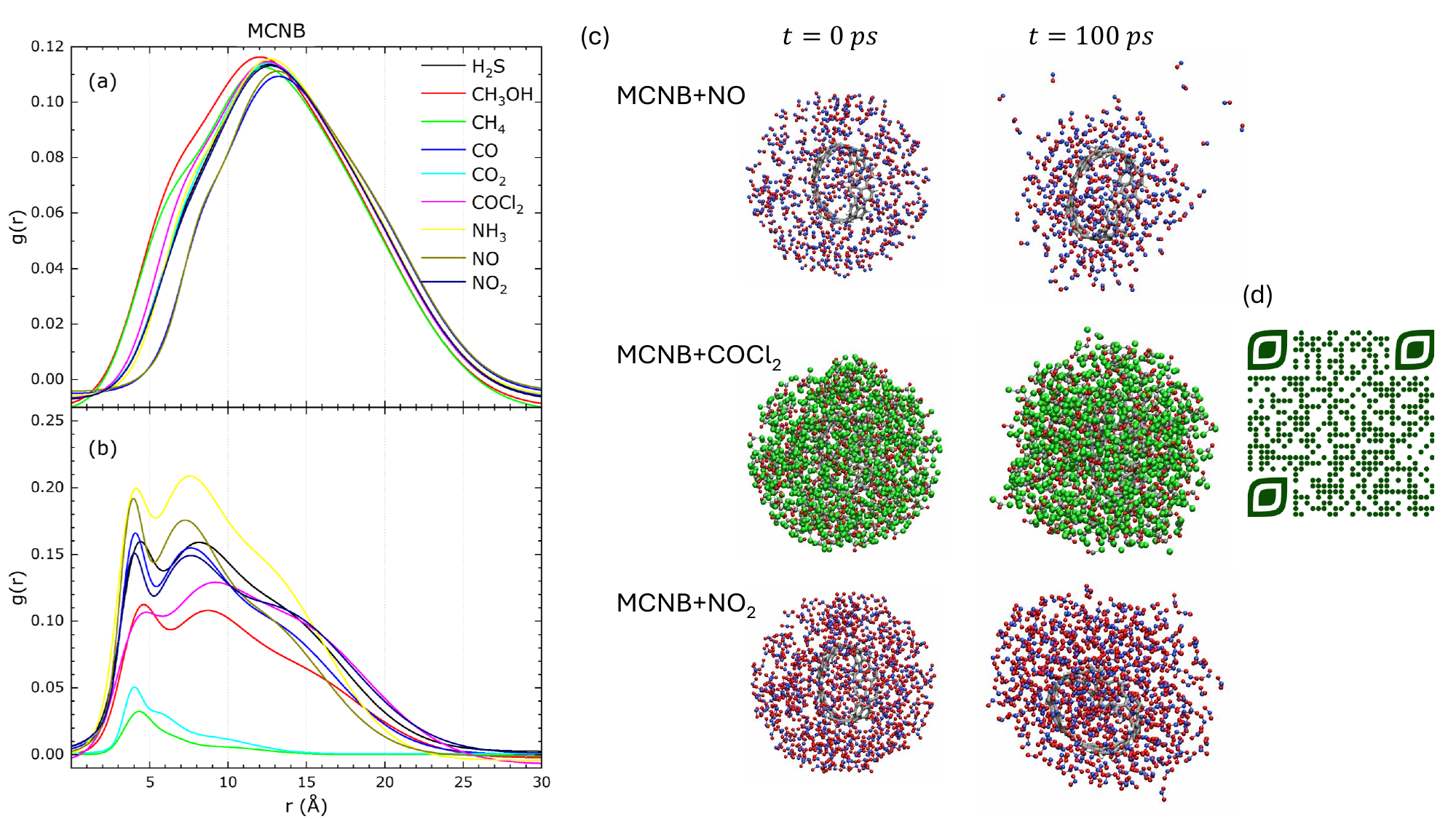}
\caption{(a) Initial, and (b), final radial distribution functions for all MCNB complexes calculated with VMD~\cite{vmd,topotools}. (c) Initial (t=0~ps) and final (t=100~ps) configurations for the best--ranked complexes. (d) QR~Code: molecular dynamics animations for nanobelts interacting with~500 gas molecules.}
\label{Fig:RDF_MCNB}
\end{figure}

The spatial distribution of molecular interactions was analyzed through radial distribution functions (RDFs) for systems containing nanobelts interacting with 500 gas molecules. Figures~\ref{Fig:RDF_CNB} and~\ref{Fig:RDF_MCNB} present the RDF analyses, with panels~\ref{Fig:RDF_CNB}(a) and~\ref{Fig:RDF_MCNB}(a) depicting the initial configurations generated using PACKMOL software. These initial distributions exhibit characteristic trimodal Gaussian profiles, reflecting the stochastic molecular placement algorithm employed by PACKMOL~\cite{packmol_0,packmol_1}. Panels~\ref{Fig:RDF_CNB}(b) and~\ref{Fig:RDF_MCNB}(b) illustrate the evolved RDFs after 100~ps of simulation time, maintaining the distinctive three-peak structure.

To quantify the spatial organization, all RDF profiles were deconvoluted using three-component Gaussian fitting. The resulting peak positions ($r_1$, $r_2$, and $r_3$), representing the most probable shell radii for particle locations, are tabulated in Tables~\ref{Tab:ResultsMDCNB} and~\ref{Tab:ResultsMDMCNB} for CNB and MCNB systems, respectively. Each table presents comparative data for both the initial configuration (t=0~ps, top row) and the equilibrated state (t=100~ps, bottom row), enabling direct assessment of the temporal evolution of molecular organization in these systems.

\begin{table}[htpb]
\caption{Shells radii, number of molecules inside a shell with 10~\AA~radius ($n$), percentage of variation on the number of molecules ($\% \Delta n$) in a shell with 10~\AA~radius after 100~ps~MD simulation time for CNB complexes$^\dagger$.}
\label{Tab:ResultsMDCNB}
\begin{center}
\renewcommand{\arraystretch}{0.7}
\begin{tabular}{lrrrrr}
  \hline
  System & $r_1$ & $r_2$ & $r_3$ & $n$ & $\Delta n$ \\
\hline
  \hline
  CNB+NO & 6.938 & 11.487 & 18.728 & 33 & ---  \\
  & 3.871 & 6.950 & 12.584 & 86 & 161  \\
\hline
  CNB+COCl\textsubscript{2}& 5.776 & 10.343 & 16.505 & 46 & ---  \\
  & 4.191 & 8.012 & 14.204 & 63 & 37  \\
\hline
  CNB+NO\textsubscript{2}  & 6.845 & 11.332 & 18.740 & 38 & ---  \\
  & 3.903 & 7.306 & 13.478 & 74 & 95  \\
\hline
   CNB+NH\textsubscript{3} & 6.209 & 10.661 & 17.457 & 42 & ---  \\
  & 3.752 & 6.956 & 12.394 & 109 & 160  \\
\hline
  CNB+CO& 7.076 & 11.525 & 18.853 & 34 & --- \\
  & 3.947 & 7.097 & 12.992 & 72 & 38  \\
\hline
  CNB+CO\textsubscript{2}& 6.803 & 11.279 & 18.488 & 37 & ---  \\
  & 3.881 & 5.162 & 8.438 & 11 & -70  \\
\hline
  CNB+CH\textsubscript{3}OH& 5.885 & 10.081 & 17.044 & 47 & ---  \\
  & 4.357 & 7.791 & 13.738 & 51 & 9  \\
\hline
  CNB+H\textsubscript{2}S & 6.717 & 11.225 & 18.649 & 38 & ---  \\
  & 4.066 & 7.390 & 13.269 & 80 & 111  \\
\hline
  CNB+CH\textsubscript{4}& 5.981 & 10.514 & 17.042 & 42 & ---  \\
  & 2.604 & 4.187 & 8.194 & 8 & -81  \\
\hline
\hline
\end{tabular}
\begin{flushleft}
\tiny {$^\dagger$ Radii are in units of \AA.}
\end{flushleft}
\end{center}
\end{table}

\begin{table}[htpb]
\caption{Shells radii, number of molecules inside a shell with 10~\AA~radius ($n$), percentage of variation on the number of molecules ($\% \Delta n$) in a shell with 10~\AA~radius after 100~ps~MD simulation time for MCNB complexes$^\dagger$.}
\label{Tab:ResultsMDMCNB}
\begin{center}
\renewcommand{\arraystretch}{0.7}
\begin{tabular}{lrrrrr}
  \hline
  System & $r_1$ & $r_2$ & $r_3$ & $n$ & $\Delta n$  \\
\hline
  \hline
  MCNB+NO & 8.018 & 11.855 & 17.663 & 25 & ---  \\
  & 3.791 & 6.649 & 11.611 & 79 & 216  \\
\hline
  MCNB+COCl\textsubscript{2}& 6.650 & 10.989 & 16.463 & 38 & ---  \\
  & 4.059 & 7.712 & 13.635 & 61 & 23  \\
\hline
  MCNB+NO\textsubscript{2}  & 6.723 & 11.180 & 16.923 & 35 & ---  \\
  & 3.840 & 6.810 & 12.777 & 68 & 94  \\
\hline
   MCNB+NH\textsubscript{3} & 7.037 & 11.242 & 16.770 & 35 & ---  \\
  & 3.785 & 6.709 & 11.768 & 98 & 180  \\
\hline
  MCNB+CO\textsubscript{2}& 6.892 & 11.105 & 16.830 & 36 & ---  \\
  & 3.899 & 5.265 & 8.362 & 10 & -72  \\
\hline
  MCNB+CH\textsubscript{3}OH& 5.820 & 10.428 & 15.805 & 45 & ---  \\
  & 4.256 & 7.806 & 13.407 & 52 & 16  \\
\hline
  MCNB+H\textsubscript{2}S& 6.670 & 10.940 & 16.406 & 35 & ---  \\
  & 4.073 & 7.220 & 12.593 & 73 & 109  \\
\hline
  MCNB+CO & 8.008 & 11.800 & 17.519 & 26 & ---  \\
  & 3.911 & 6.900 & 12.169 & 71 & 173  \\
\hline
  MCNB+CH\textsubscript{4}& 5.614 & 10.537 & 15.840 & 43 & --- \\
  & 4.138 & 5.725 & 9.228 & 6 & -86  \\
\hline
\hline
\end{tabular}
\begin{flushleft}
\tiny {$^\dagger$ Radii are in units of \AA.}
\end{flushleft}
\end{center}
\end{table}

Analysis of the shell radii revealed a systematic decrease across all measured values, suggesting that both nanobelt architectures function as attractive potential wells for the investigated gas species. To quantitatively characterize this attraction phenomenon, we evaluated the molecular population dynamics within a 10~\AA radius from the nanobelt surface. This was accomplished by comparing the molecular count before ($n_b$) and after ($n_a$) the simulation period, expressed as $\Delta n=n_a - n_b$.

The sign of $\Delta n$ serves as a critical parameter in determining the dominant interaction regime: positive values indicate that the nanobelt-gas attractive forces predominate, while negative values suggest that intermolecular repulsion between gas molecules exceeds the nanobelt attraction potential. Our analysis revealed a distinct behavioral dichotomy: CH\textsubscript{4} and CO\textsubscript{2} exhibited negative $\Delta n$ values, while all other investigated gases demonstrated positive values.

A significant consideration in interpreting these results is the molecular volume effect. Despite showing favorable adsorption energetics (as documented in Tables~\ref{Tab:ResultsAdsCNB} and~\ref{Tab:ResultsAdsMCNB}), larger molecules such as COCl\textsubscript{2} and CH\textsubscript{3}OH displayed relatively modest positive $\Delta n$ values. This observation suggests that steric effects play a crucial role in determining the ultimate molecular population density near the nanobelt surface, even when electronic interactions are favorable.

\begin{figure}[htpb]
\centering
\includegraphics[width=14cm]{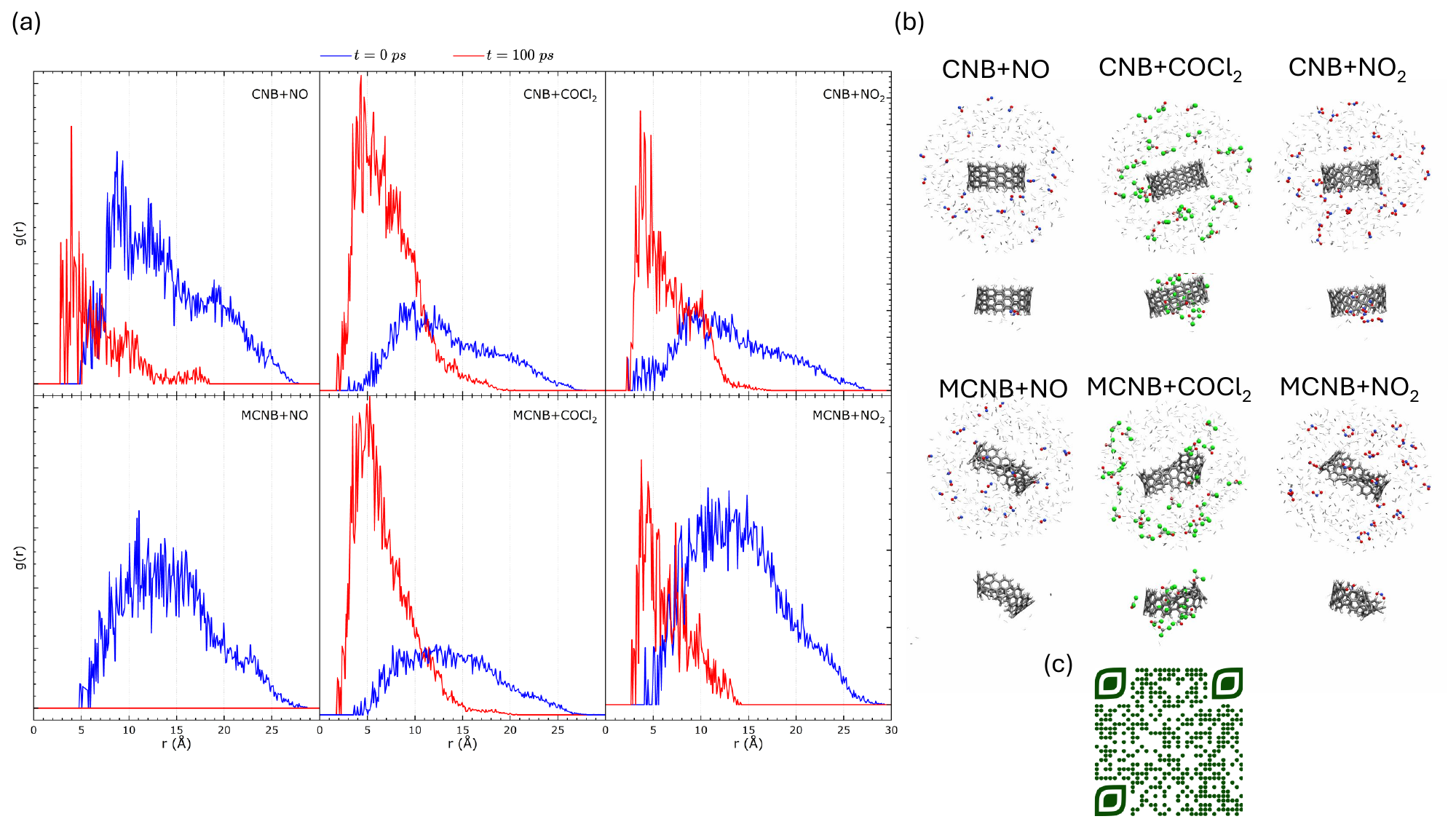}
\caption{(a) Initial (solid blue line) and final (solid red line) radial distribution functions for both nanobelts best--ranked complexes calculated with VMD~\cite{vmd,topotools}. (b) Initial and final configurations for the best--ranked complexes (t=0~ps, top row; $t=100\,ps$, bottom row; respectively). (c) QR Code: molecular dynamics animations for nanobelts interacting with the air mixture.}
\label{Fig:RDF_AirMix}
\end{figure}

To better approximate real-world conditions, we constructed a more complex simulation environment using PACKMOL. The model consisted of a dry air matrix containing~500 molecules, doped with 5\% (by volume) of the highest-affinity greenhouse gases identified in our previous analyses. The temporal evolution of molecular distributions was analyzed through radial distribution functions (RDFs) at both initial (t=0~ps) and final (t=100~ps) states, with corresponding spatial configurations presented in Figure~\ref{Fig:RDF_AirMix}.

Analysis of the RDF profiles (Figure~\ref{Fig:RDF_AirMix}(a)) revealed a systematic shift toward smaller radial distances for all systems except MCNB+NO, indicating the presence of attractive interactions between the nanobelt structures and the greenhouse gas molecules. This behavior was observed consistently for both CNB and MCNB architectures. The spatial configurations depicted in Figure~\ref{Fig:RDF_AirMix}(b) (initial state: top row; final state: bottom row) demonstrate preferential accumulation of greenhouse gas molecules in proximity to the nanobelt surfaces, accompanied by the displacement of atmospheric components (N\textsubscript{2}, O\textsubscript{2}, and Ar) from the immediate vicinity of the nanostructures. This selective molecular segregation suggests a promising potential for greenhouse gas separation applications.

The molecular dynamics simulations of both densely packed systems and dry air mixtures provide compelling evidence for the efficacy of CNB and MCNB architectures as potential greenhouse gas capture platforms. The observed selective molecular accumulation and sustained binding interactions in both simulation environments strongly support the viability of these nanostructures for practical gas separation and sequestration applications. These findings are particularly significant given that the simulations encompassed realistic environmental conditions, including competitive binding scenarios and complex molecular matrices. The demonstrated molecular selectivity and capture stability suggest that these nanobelt systems could serve as promising candidates for next-generation greenhouse gas mitigation technologies.

\section{CONCLUSIONS}
\label{Sec:Conclusions}

Our computational analysis revealed thermodynamically favorable interactions between both carbon nanobelts (CNB and MCNB) and the investigated greenhouse gases, demonstrated by negative adsorption energies across all studied systems. The Möbius variant (MCNB) exhibited consistently stronger binding affinities compared to the unmodified CNB, with both structures showing particularly robust interactions with NO, COCl\textsubscript{2}, and NO\textsubscript{2}, as detailed in Tables~\ref{Tab:ResultsAdsCNB} and~\ref{Tab:ResultsAdsMCNB}.

Electronic structure analysis through HOMO/LUMO surface calculations indicated negligible perturbation upon gas adsorption, confirming the electronic stability of both nanobelt configurations. Topological bond analysis provided detailed insights into the interaction mechanisms: CNB established one covalent bond with NO, seven non-covalent interactions with COCl\textsubscript{2}, and two non-covalent bonds with NO\textsubscript{2}. The Möbius nanobelt (MCNB) demonstrated enhanced binding characteristics, forming two covalent and one non-covalent bonds with NO, ten non-covalent interactions with COCl\textsubscript{2}, and five non-covalent bonds with NO\textsubscript{2}.

The interaction mechanisms were classified as chemisorption for NO, COCl\textsubscript{2} and NO\textsubscript{2}, while physisorption dominated the interactions with other greenhouse gases. Temporal analysis revealed sub-millisecond recovery times for most systems, indicating excellent reusability potential. The notable exception was the MCNB+NO complex, where covalent bonding was observed. Despite the strong binding characteristics, both nanobelt variants exhibited minimal changes in electrical conductivity upon gas adsorption, as evidenced by the calculated sensitivity factors.

Molecular dynamics simulations of mixed-gas environments, including packed and dry air conditions, corroborated the greenhouse gas capture capabilities of both nanobelts. This was quantitatively demonstrated through the observed shifts in radial distribution function peak positions and comparative analysis of initial versus final particle distributions, validating their potential application in gas capture technologies.

\section*{CRediT AUTHORSHIP CONTRIBUTION STATEMENT}
\textbf{C. Aguiar}: Investigation, Formal analysis, Writing-original draft, Writing-review \& editing.
\\
\textbf{I. Camps}: Conceptualization, Methodology, Software, Formal analysis, Resources, Writing-review \& editing, Supervision, Project administration.

\section*{DECLARATION OF COMPETING INTEREST}

The authors declare that they have no known competing financial interests or personal relationships that could have appeared to influence the work reported in this paper.

\section*{DATA AVAILABILITY}
The raw data required to reproduce these findings are available to download from \href{https://doi.org/10.5281/zenodo.14010167}{https://doi.org/10.5281/zenodo.14010167}.

\section*{ACKNOWLEDGEMENTS}
We would like to acknowledge financial support from the Brazilian agencies CNPq, CAPES and FAPEMIG. Part of the results presented here were developed with the help of CENAPAD-SP (Centro Nacional de Processamento de Alto Desempenho em S\~ao Paulo) grant UNICAMP/FINEP-MCT, and the National Laboratory for Scientific Computing (LNCC/MCTI, Brazil) for providing HPC resources of the Santos Dumont supercomputer.

\newpage
\bibliographystyle{elsarticle-num}
\bibliography{unifal}

\end{document}